\documentclass[conference]{IEEEtran}
\usepackage{cite}
\usepackage{amsmath,amssymb,amsfonts,gensymb}
\usepackage{algorithmic}
\usepackage{graphicx}
\usepackage{textcomp}
\usepackage{xcolor}
\usepackage{fancyhdr}
\usepackage[hyphens]{url}

\usepackage{subfigure}
\usepackage{enumitem}
\usepackage{array,booktabs}
\usepackage{makecell}
\usepackage[mathscr]{euscript}
\usepackage{xspace}

\usepackage{float}
\usepackage{hyperref}
\usepackage{multirow}
\DeclareMathSizes{6}{6}{4}{3}

\def\spade{\textsc{Spade}\xspace}
\def\otfspade{\textsc{Otf-Spade}\xspace}
\def\spadex{\textsc{Spade}}
\def\soar{\textsc{Soar}\xspace}
\def\carom{\textsc{Carom}\xspace}
\def\soarx{\textsc{Soar}}
\def\aos{\textsc{Accelerator for Spatially Sparse 3D DNN}s\xspace}

\def\BibTeX{{\rm B\kern-.05em{\sc i\kern-.025em b}\kern-.08em
    T\kern-.1667em\lower.7ex\hbox{E}\kern-.125emX}}
\def\spnna{\textsc{Ss}p\textsc{nna}\xspace}
\def\admac{Ad\textsc{mac}\xspace}
\def\coir{\textit{\textsc{Coir}}\xspace}
\newcommand{\bfit}[1]{\textbf{\textit{#1}}}
\newcommand{\unit}[1]{\underline{\textit{#1}}}

\newcommand{\redc}[1]{}

\def\etal{\textit{et al. }}

\graphicspath{{./images/}}

\usepackage{mathtools}
\DeclarePairedDelimiter{\ceil}{\lceil}{\rceil}

\fancypagestyle{firstpage}{
  \fancyhf{}

  \fancyfoot[C]{\thepage}
}

\title{AccSS3D: Accelerator for Spatially Sparse 3D DNNs}

\author{
  \IEEEauthorblockN{Om Ji Omer\IEEEauthorrefmark{1}\IEEEauthorrefmark{4}, Prashant Laddha\IEEEauthorrefmark{1}, Gurpreet S Kalsi\IEEEauthorrefmark{1},
          Anirud Thyagharajan\IEEEauthorrefmark{1}, Kamlesh R Pillai\IEEEauthorrefmark{1},\\
          Abhimanyu Kulkarni\IEEEauthorrefmark{2},
                  Anbang Yao\IEEEauthorrefmark{3}, Yurong Chen\IEEEauthorrefmark{3}, Sreenivas Subramoney\IEEEauthorrefmark{1}}
                  \IEEEauthorblockA{\IEEEauthorrefmark{1}Processor Architecture Research Lab, Intel Labs, Bangalore}
                  \IEEEauthorblockA{\IEEEauthorrefmark{2}University of Wisconsin - Madison, work done while at Intel}
                  \IEEEauthorblockA{\IEEEauthorrefmark{3}Intel Labs, China}
                  \IEEEauthorblockA{\IEEEauthorrefmark{4}Corresponding Author : Om Ji Omer, om.j.omer@intel.com}
                }

\begin{document}
\maketitle
\thispagestyle{firstpage}
\pagestyle{plain}

\def\convspsc{\textbf{36.6x}\xspace}
\def\convspfc{\textbf{16.8x}\xspace}
\def\nettspsc{\textbf{23.7x}\xspace}
\def\nettspfc{\textbf{11.8x}\xspace}
\def\convessc{\textbf{2079x}\xspace}
\def\convesfc{\textbf{2232x}\xspace}
\def\nettessc{\textbf{23.2x}\xspace}
\def\nettesfc{\textbf{24.8x}\xspace}
\def\nettesscdram{\textbf{18.6x}\xspace}
\def\nettesfcdram{\textbf{19.5x}\xspace}
\def\convprsc{\textbf{56.8x}\xspace}
\def\convprfc{\textbf{132.9x}\xspace}
\def\corearea{\textbf{0.92 mm2}\xspace}

\begin{abstract}
Semantic understanding and completion of real-world scenes is a foundational primitive of 3D Visual perception widely used in high-level applications such as robotics, medical imaging, autonomous driving and navigation. Due to the curse of dimensionality, compute and memory requirements for 3D scene understanding grow in cubic complexity with voxel resolution, posing a huge impediment to realizing real-time energy-efficient deployments. The inherent spatial sparsity present in the 3D world due to free space is fundamentally different from the channel-wise sparsity that has been extensively studied. We present \aos (AccSS3D) , the first end-to-end solution for accelerating 3D scene understanding by exploiting the ample spatial sparsity. As an algorithm-dataflow-architecture co-designed system specialized for spatially-sparse 3D scene understanding, AccSS3D includes novel spatial locality-aware metadata structures, a near-zero latency and spatial sparsity-aware dataflow optimizer, a surface orientation aware pointcloud reordering algorithm and a co-designed hardware accelerator for spatial sparsity that exploits data reuse through systolic and multicast interconnects. The \spnna accelerator core together with the 64 KB of L1 memory requires \corearea of area in 16nm process at 1 GHz. Overall, AccSS3D achieves \convspfc speedup and a \convesfc energy efficiency improvement for 3D sparse convolution compared to an Intel-i7-8700K 4-core CPU, which translates to a \nettspfc end-to-end 3D semantic segmentation speedup and a \nettesfc energy efficiency improvement (iso technology node).
\end{abstract}

\section{Introduction}

Understanding 3D geometry and semantics of a scene is essential to many real-world systems including but not limited to autonomous driving, robotics, remote sensing, AR/VR and medical treatment \cite{liu2018see, li2019rgbd, garbade2019two, zhang2019cascaded, chen2017multi}.
With rapid growth in 3D acquisition technologies, 3D sensors are becoming increasingly available and affordable for rich data generation through various types of 3D scanners, LiDARs and RGB-D cameras.%
3D data is usually represented in various formats like pointclouds, meshes, depth maps and volumetric grids.
Deep Learning techniques have disrupted traditional methods in domains such as computer vision, speech processing and machine translation that operate over images, videos, audio, text and other forms of data.
However, DL methods face severe challenges in processing 3D visual data due to the high dimensionality and the unstructured nature of 3D data.

\begin{figure}
	\centering
	\includegraphics[trim=15 20 15 18, clip, width=0.5\textwidth]{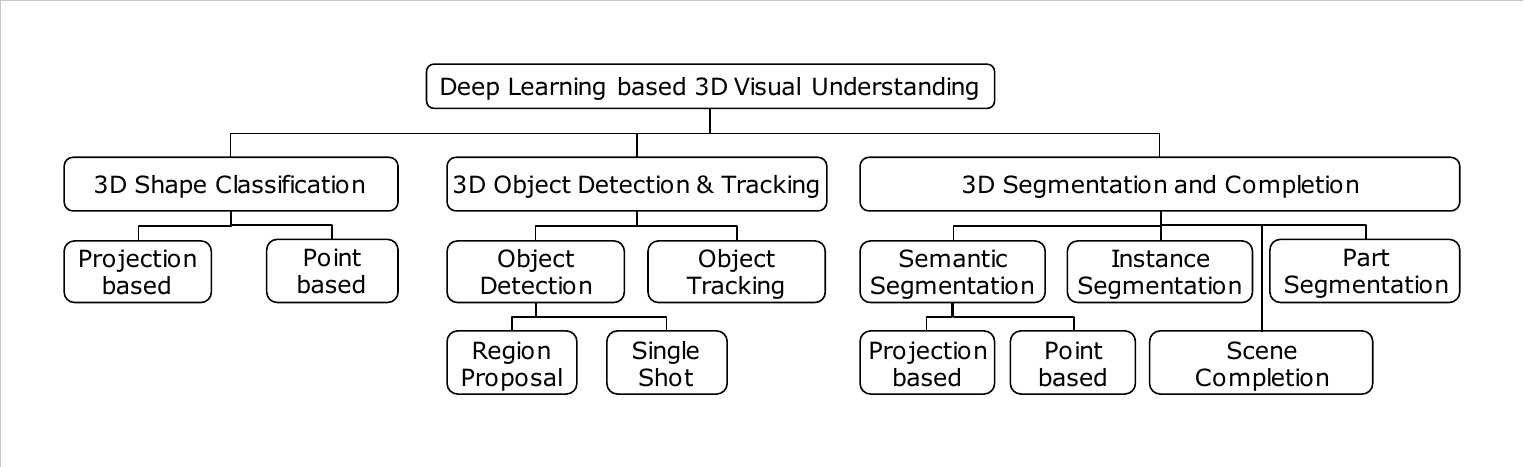}
	\caption{Taxonomy of 3D visual data processing, courtesy: \cite{guo2019deep}}
        \label{fig:taxonomy_3d_proc}
\end{figure}

\begin{figure}%
	\centering
	\includegraphics[width=0.4\textwidth]{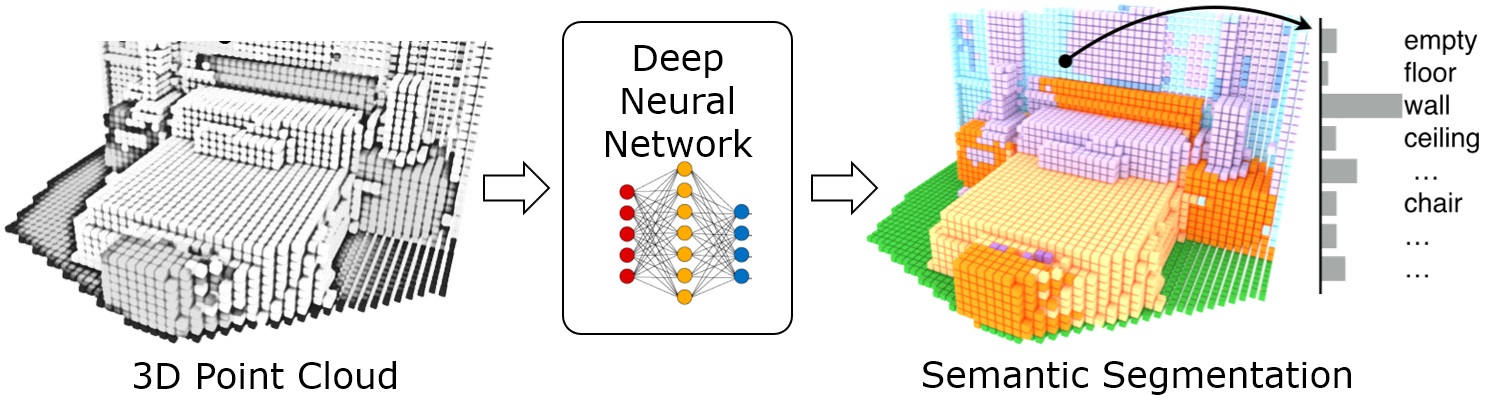}
    \caption{3D Semantic segmentation, annotating voxels with labels \cite{song2017semantic}}%
        \label{fig:scene_semantics_completion}
\end{figure}

Several methods (Figure \ref{fig:taxonomy_3d_proc}) have been proposed for various 3D visual AI applications such as shape classification, object detection, tracking, and scene segmentation.
3D scene segmentation requires extraction of global geometric structure and intricate details of each point in the 3D pointcloud.
Semantic segmentation (Figure \ref{fig:scene_semantics_completion}) aims to divide a pointcloud into several subsets based on real world semantics. Instance segmentation identifies instances of each semantic object, while part segmentation further breaks each instance into different parts. Scene completion predicts and labels missing regions arising from sensing inaccuracies or occlusions.

Point based networks (Table \ref{tab:seg_methods_mIoU}) directly operate over irregular, orderless and unstructured pointclouds, making it infeasible to directly apply standard CNNs. Performance of multi-view 2D projection based methods is highly sensitive to viewpoint selection and these methods also do not fully exploit underlying 3D geometry.  
Since volumetric representation preserves neighbourhood structure of pointcloud and regular data formats allow direct application of standard 3D convolutions, volumetric projection based methods provide higher accuracy. %
However, voxelization introduces discretization artifacts and causes information loss. 
A low resolution voxel representation can degrade accuracy, whereas high resolution pointclouds cubically grow compute and memory requirements (Figure \ref{fig:accu_vs_resolution}).
This poses a huge impediment in achieving real-time real-world deployments of visual AI that fundamentally require low-latency and low-energy 3D scene understanding capability.
Thankfully, the ample free space in 3D scenes provides significant opportunity to exploit the inherent \textbf{spatial sparsity}.

\begin{table}
  \centering
  \caption{Comparison of semantic segmentation methods on ScanNet\cite{dai2017scannet}}
  \label{tab:seg_methods_mIoU}
  \resizebox{0.5\textwidth}{!}{
    \begin{tabular}{c|c|c|c|c|c|c|l}
      \hline
      \multicolumn{2}{c|}{\textbf{Category}} & \textbf{Method} & \textbf{mIoU} & \multicolumn{2}{c|}{\textbf{Category}} & \textbf{Method} & \textbf{mIoU} \\ \hline \hline
      \multirow{5}{0.2cm}{\rotatebox[origin=c]{90}{Point based}} & \multirow{2}{1.5cm}{Pointwise MLP}    & PointNet++ \cite{qi2017pointnet++}  & 33.9 & \multirow{5}{0.5cm}{\rotatebox[origin=tc]{90}{\makecell[b]{Projection\\based}}} & \multirow{2}{*}{\textbf{Volumetric $\star$}} & SparseConvNet \cite{graham20183d}  & \textbf{72.5 $\star$} \\
      \cline{3-4}\cline{7-8}          &                                        & PointSIFT \cite{jiang2018pointsift}  & 41.5 &            &                             & MinkowskiNet \cite{choy20194d}   & \textbf{73.6 $\star$} \\
      \cline{2-4}\cline{6-8}          & \multirow{2}{1.5cm}{Point Convolution} & PointCNN \cite{li2018pointcnn}    & 45.8 &            & Multiview                   & TangentConv \cite{tatarchenko2018tangent}    & 40.9 \\
      \cline{3-4}\cline{6-8}          &                                        & KPConv \cite{thomas2019kpconv}      & 68.4 &            & Perm. Lattice               & LatticeNet \cite{rosu2019latticenet}     & 64.0 \\
      \cline{2-4}\cline{6-8}          & Graph based                            & HPEIN \cite{jiang2019hierarchical}      & 61.8 &            & Hybrid                      & UPB \cite{chiang2019unified}            & 63.4 \\ \hline
    \end{tabular}
  }
\end{table}

For dense DNNs, multiple accelerator designs \cite{zhang2016cambricon, parashar2017scnn, gondimalla2019sparten, hegde2019extensor, centaur} have been proposed to exploit \textit{channel-wise sparsity} in weights and inputs.
They focus on identifying valid pairs of non-zero weights and input activations while operating over a compressed data structure.
We observe two fundamental attributes of spatial sparsity as present in 3D visual analytics that differentiate it from the channel-wise sparsity:
1) the selection of a non-zero pair of weights and inputs depends solely on local 3D spatial sparsity, 2) the nature of the operation per selected pair is of the matrix-to-vector type as compared to scalar-to-scalar type in channel-wise sparsity.
The first motivates a different hardware-architecture to enable input-weight pair selection based on the variable local sparsity per voxel.
The second further offers the opportunity to exploit higher efficiency through coarser dispatches of work to the compute units quite unlike the prior accelerators that hunt for fine-grained sparsity across channels.
\bfit{We will show in Section \ref{sec:sec_motiv} that a custom-built sparse engine is required for efficient acceleration of spatially-sparse applications such as 3D visual analytics as they have distinct hardware requirements. }

Optimized dataflows, as explored in prior art \cite{chen2016eyeriss, chen2018eyeriss, yang2016systematic, hegde2018morph, gao2019tangram, kwon2018mef}, focus on dense-DNNs and enable tiled execution with optimal tile-sizes and loop-orders to exploit data-locality in on-die memories.
It is worth noting that the execution attributes of dense-DNNs (\#data transfers and \#compute operations) are highly predictable, and since they do not depend on input data, optimal dataflows can be derived apriori based on network parameters and underlying hardware architecture.
However, these optimizers cannot be deployed in data-dependent sparse-DNN contexts since determining optimal tile-size and loop order requires on-the-fly processing of input data and hence cannot be determined a priori.
Further, due to the sparsity-induced asymmetry of work packets, load balancing requires online intelligence as part of a suitably designed software-hardware solution.
\bfit{Hence specialized dataflow optimizers that explicitly exploit spatial sparsity structure are warranted.
In addition, they must achieve compute efficiencies similar to dense-DNNs without the latency overheads incurred by on-the-fly processing. }

\subsection{Our Contribution}
Driven by a detailed end-to-end performance characterization of the Sparse Convolution Network (SCN), we present \textbf{\aos}, the first end-to-end solution for accelerating 3D scene segmentation by exploiting spatial sparsity.
In brief, our key contributions are as follows:\newline
  1. We introduce a \bfit{novel locality-aware metadata structure}, \coir, for voxel data that stores spatial locality information by encoding active receptive fields at each location. We also present a \bfit{novel pointcloud reordering technique}, \soar, that maximizes spatial data locality and reuse across all DNN layers for multi-level memory architectures.\newline %
  2. We present (to the best of our knowledge) the \bfit{first-ever sparsity aware dataflow optimizer}, \spade, that analyzes the local
  geometry structure of a pointcloud and uses the sparsity attributes to maximize data reuse across all DNN layers. To meet real-time requirements, we enable latency-critical dataflow optimizer routines to be run in offline mode while retaining the benefits of the sparsity-aware dataflow. \newline %
  3. We present a \bfit{novel sparse-accelerator for spatially-sparse DNNs}, \spnna, whose sparse scheduler maximizes weight reuse. Systolic and multicast interconnects in the PE arrays maximize input feature-map reuse. We also propose, \bfit{\admac}, an accelerator for fast 3D neighbourhood search and \coir metadata creation and \bfit{\carom}, a joint dataflow optimization technique for multi-level memory hierarchies. We show how a scaled-up architecture based on AccSS3D accelerates 3D sparse convolution by \convspfc compared to a 4-core CPU baseline with a \convesfc improvement in energy efficiency.\newline %

\section{Background}
Recent algorithmic advances \cite{dai2018scancomplete, wu20153d, maturana2015voxnet} extended 2D-CNNs to 3D to recognize 3D objects and enabled semantic segmentation of real-world 3D scenes.
However, due to the curse of dimensionality, compute and memory requirements grow in cubic complexity with voxel resolution.

Free space in 3D scenes is a source of inherent spatial sparsity, offering high opportunity for efficient processing. 
Recent algorithms \cite{qi2017pointnet, qi2017pointnet++, graham20183d} exploit this spatial sparsity by processing only those voxels (\textit{active voxels}) that are in the vicinity of a surface boundary.
They transform 3D sparse data into a list of active voxels for compressed storage and use a map \cite{googlehashmap, teschner2003optimized} to retrieve lists of indices based on 3D coordinates.
Graham \etal \cite{graham2017submanifold} introduced \textit{Sparse Convolution Network (SCN)} using the novel \textit{Valid Sparse Convolution (VSC)} that provides significant reduction in compute and memory costs by considering output voxels as active only if the corresponding input voxel is active.
\textit{This retains the sparsity structure of feature maps across the network and paves the way for achieving real-time complex 3D scene understanding.}
Zhang \etal \cite{zhang2018efficient} extended SCN to use Spatial Group Convolution, dividing the pointcloud into groups, further reducing compute without accuracy loss.
Several systems employ encoder-decoder U-net topologies \cite{shi2019pv, wang20203d, dai2019sg, ye2019arpnet, li2019pointsite, wu2019complementary, yan2018second, schmohl2019submanifold, di2019wafer, ye2020sarpnet}, with layers that reduce spatial resolution (strided convolutions) followed by upsampling layers (deconvolutions), interspersed with VSC layers.
\begin{figure}
  \centering
  \includegraphics[width=0.4\textwidth]{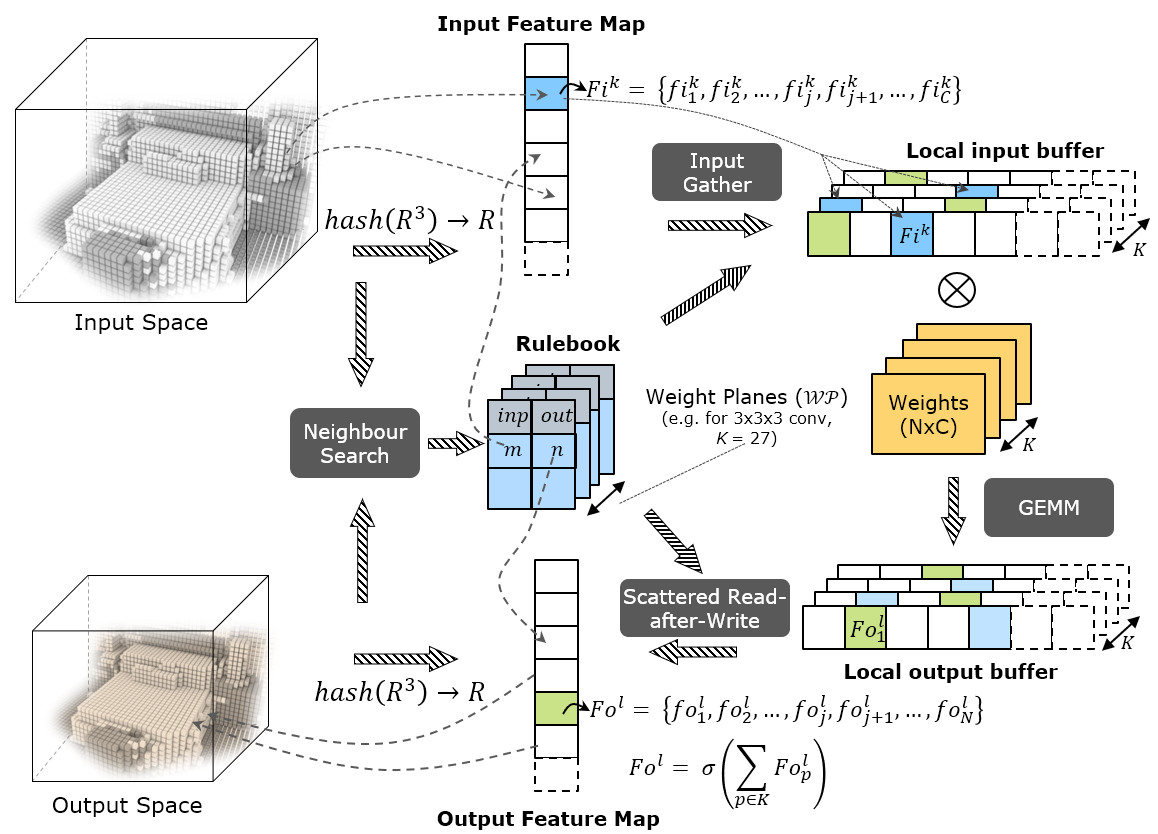}
  \caption{Processing flow for SCN workload}
  \label{fig:scn_wl_proc}
\end{figure}
\begin{figure*}[t]
  \centering
  \includegraphics[width=0.95\textwidth]{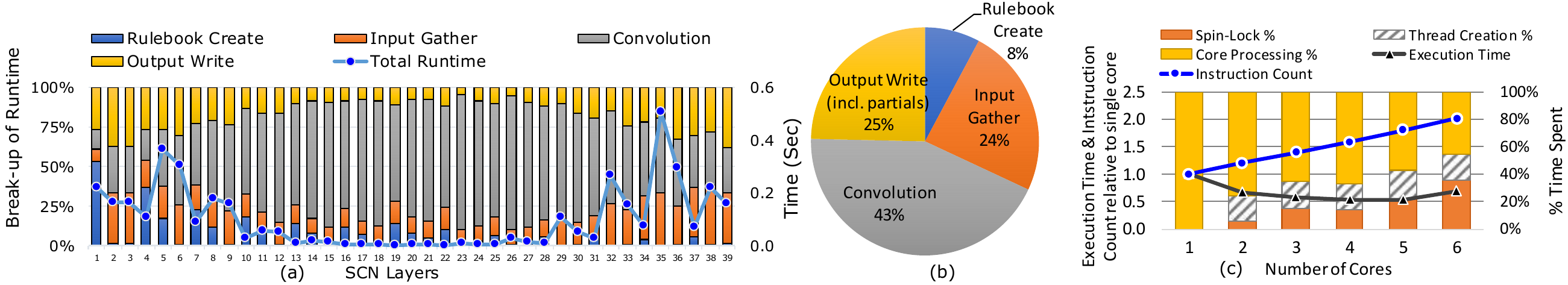}
  \caption{Performance characterization of SCN\cite{graham20183d} inference for a typical ScanNet \cite{dai2017scannet} dataset on Intel-i7-8700K CPU with single core execution @ 3.7GHz (a) Layer-wise runtime and break-up into constituent functions (b) Runtime break-up across all layers (c) Performance scaling with multi-core execution}
  \label{fig:wl_profile_breakup}
\vspace{-0.8cm}
\end{figure*}

\section{Motivation}
\label{sec:sec_motiv}
\subsection{Increasing complexity of 3D Visual processing}
Over the years, compute/memory requirements for 3D visual processing have increased multi-fold. (Table \ref{tab:dataset_size}). %
While 3D volumetric representations enable higher accuracy (Table \ref{tab:seg_methods_mIoU}), low-resolution representation can hurt accuracy (Figure 5). Though compute and memory requirements grow in cubic order with resolution size, the free space in 3D scenes, a source of exploitable ’spatial sparsity’, motivates a hardware-software co-designed approach to acceleration of 3D visual processing.
\begin{figure}
	\centering
	\includegraphics[width=0.40\textwidth]{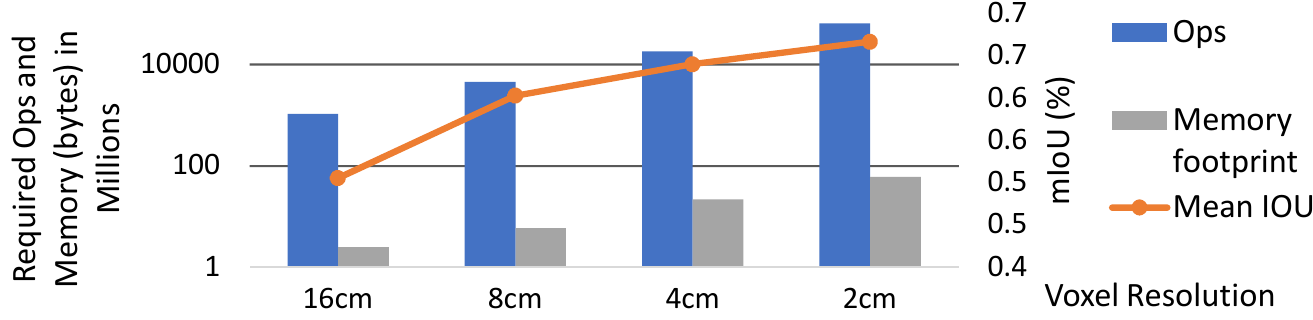}
	\caption{Correlation between compute needs and accuracy (mIoU) as a function of resolution obtained by downsampling input pointclouds in ScanNet\cite{dai2017scannet}.}
        \label{fig:accu_vs_resolution}
\end{figure}

\begin{table} [b]
  \centering
  \caption{A chronological overview of typical RGB-D/3D datasets.}
  \label{tab:dataset_size}
  \resizebox{0.5\textwidth}{!}{
  \begin{tabular}{cccc}
    \toprule
    \textbf{Datasets} & \textbf{\# Images/Models (\#Objects)} & \textbf{\#Ops} \cite{graham20183d} & \textbf{I/O FP (grid)} \cite{graham20183d} \\
    \midrule
    NYU Depth V2 \cite{Silberman_ECCV12} & 1.5k images (894) & $\sim$2.3B & $\sim$12MB ($4096^3$)\\
    ModelNet \cite{Wu_2015_CVPR} & 50k models (660) & $\sim$9M & $\sim$400kB ($30^3$)\\
    ShapeNet \cite{chang2015shapenet} & 3M models (3135) & $\sim$0.5B & $\sim$1MB ($192^3$)\\
    Matterport3D \cite{chang2017matterport3d} & 194k images (40) & - & -\\
    ScanNet \cite{dai2017scannet} & 2.5M frames (40) & $\sim$90B & $\sim$65MB ($4096^3$)\\
    Waymo \cite{sun2019scalability} & 113K models (210K) & $\sim$249B & $\sim$128MB ($4096^3$)\\
    \bottomrule
  \end{tabular}
}
\end{table}

\subsection{Sparse Convolution Network (SCN) workload profile}
\label{sec:wl_char}
Figure \ref{fig:wl_profile_breakup} shows the runtime CPU profile of SCN \cite{graham20183d} as a representative workload. The middle layers take much lower execution time than initial and last few layers as in the U-net, the middle layers operate on lowest resolution needing lower compute.
As the order of voxel processing can be different from the memory layout, a gather operation for inputs and scattered write for outputs is required.
Overall, \textit{Input Gather and Output Write dominates across the layers, due to the weight stationary dataflow, where input-output index pairs are created and processed for each weight plane independently.}
These components are majorly dominant for the initial and last few layers, where the voxel resolution sizes are higher.
Lack of tiled execution and inability of CPU caches to hold the entire large pointcloud significantly hurts data reuse.

\begin{figure}
  \centering
  \includegraphics[width=0.50\textwidth]{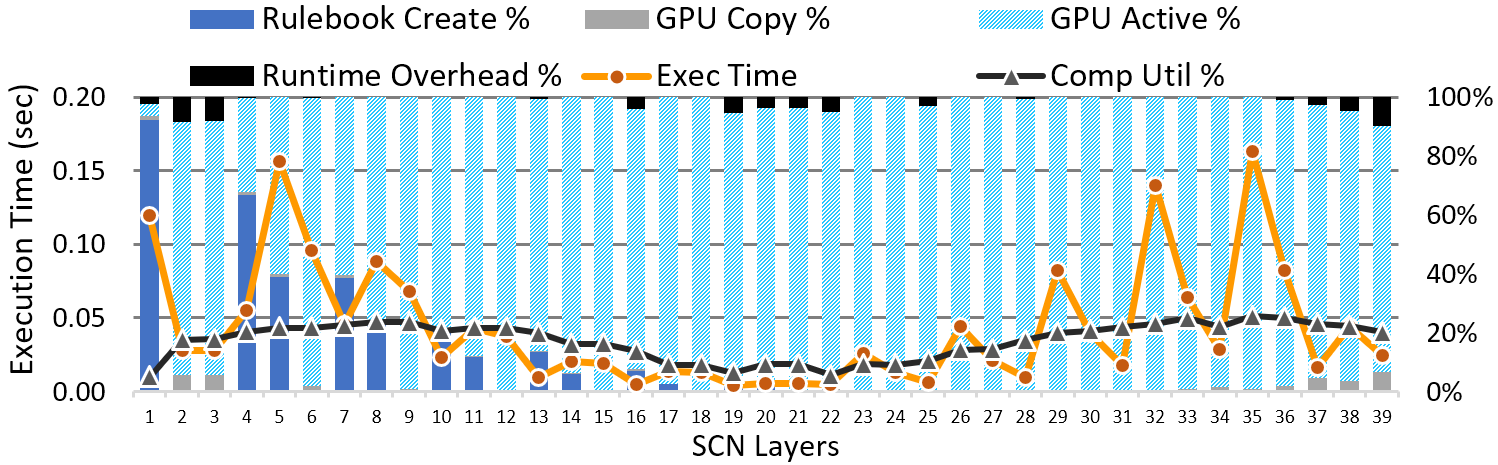}
  \caption{SCN performance characterization on Nvidia GeForce-GTX-1080 @1.6GHz. Though GPU is processing for most of the time, average GPU core utilization is significantly low at 21\% due to high register usage.}
  \label{fig:gpu_wl_profile}
\end{figure}

We observe that the performance of the workload scales poorly with number of cores and flattens beyond 4 cores (Figure \ref{fig:wl_profile_breakup}-c). 
Further analysis shows that thread synchronization (spin locks) and thread creation events (both of which scale with \#cores), increased load/store traffic for shared data, and increased DRAM bandwidth bottlenecks are responsible for this lack of scaling.
On an Nvidia GPU, SCN performance improves by 2.34X over single-core CPU (Figure \ref{fig:gpu_wl_profile}), but effective GPU utilization is low due to high register usage per thread, thereby limiting the number of threads that simultaneously execute. By recompiling GPU kernels to enforce fewer registers per thread, running thread count can be increased, but performance saturates due to memory load-store bottlenecks.

\subsection{The need for Sparsity-aware Dataflow Optimizers}
\label{sec:need_for_df}
Several dataflow optimizers have been proposed for dense DNNs \cite{chen2016eyeriss, chen2018eyeriss, yang2016systematic, hegde2018morph, gao2019tangram, kwon2018mef}.
Since dense DNNs operate over regular data-structures, memory size requirements and data accesses for each data-type can be estimated and an optimal dataflow could be chosen offline without requiring input data.
Since network parameters are usually known prior to input data arrival, an optimal dataflow could be chosen in offline mode per layer.
\bfit{Since 3D scenes are typically spatially sparse, a uniform 3D tiling strategy would result in extremely inefficient execution due to excessive memory consumption and uneven work distribution.}
Though tiling 1D compressed data-structure through a spatial hash map is a concise representation, it has the following drawbacks: 1) size of compressed data-structure varies per input pointcloud and across different regions within a pointcloud 2) memory requirement and data accesses can not be formulated as mathematical expressions 3) storing 3D data in an unordered 1-D compressed format results in irregular data accesses as convolution operations need to be performed on spatially proximate points in 3D space.
\textit{In Sections \ref{sec:coir}, \ref{sec:soar}, \ref{sec:spade_df}, we show how specialized sparsity-aware dataflow optimizers can effectively addresses these challenges.}

\subsection{Accelerator Options for 3D Spatial Sparsity}
As discussed in Section \ref{sec:need_for_df}, storing 3D spatially-sparse data in 1D compressed format leads
to inefficient execution due to irregular data accesses as convolution operations need to be performed on spatially proximate points in 3D space while the data is stored in 1D compressed format.
We discuss two potential ways to accelerate spatially sparse 3D processing: \newline
\bfit{1) Generic GEMM-based Acceleration:} Implement 3D sparse convolution as a dense-GEMM using efficient gather/scatter operations and performing GEMM over the regularized data structure similar to reference CPU implementation (Figure \ref{fig:scn_wl_proc}).\newline
\unit{Challenges:} a) Since a 3D convolution is mapped onto a GEMM engine, input/output features need to be re-fetched as many times as the local receptive field, requiring prohibitively large data transfers and bandwidth. b) Due to input-dependent sparsity, data-specific source/destination rulebook tables need to be created by the host to enable a gather/scatter engine to perform explicit data copies. When local sparsity is medium-to-high and the point-cloud is large, the rulebook can significantly increase size and bandwidth requirement. \newline
\bfit{2) Specialized Acceleration for 3D sparsity}: Build specialized implementation to locate input/output data on-the-fly while performing 3D sparse convolution on a pool of MACs. \newline
\unit{Challenges:} a) While custom acceleration for 3D sparsity can entirely eliminate the need for explicit metadata/rulebook generation, re-computation of input/output data addresses would be required for each DNN layer with the same resolution and at every accelerator core in the system. b) The number of lookups into the sparse hash-map required per unit compute is input sparsity-dependent. So, either the convolution engine will suffer significant utilization loss due to the high variability in the local receptive field OR to cover the variable latency and bandwidth requirement, the local buffering would need to be prohibitively large.
\textit{We propose in section \ref{sec:uarch} a semi-specialized accelerator for 3D sparsity which can process a compressed metadata structure while adopting a tile-based sparsity-aware dataflow for efficient data-movement}. 

\subsection{Custom Accelerator Requirements for 3D Spatial Sparsity}
\label{sec:need_for_acc}
State-of-the-art sparse accelerators \cite{hegde2019extensor, zhang2016cambricon, parashar2017scnn, Smash} have been proposed for sparse matrix multiply or sparse DNN models. In general, these sparse accelerators are composed of two major blocks: 1) a Front-end which processes metadata of input operands (feature-maps and weights) locating pairs of non-zero input operands and 2) a Back-end for performing valid operations (MAC, activation, pooling etc.) through PEs. Proposals such as Sparten \cite{gondimalla2019sparten} also schedule work efficiently such that the imbalance between compute engines can be mitigated.
\begin{table} [b]
  \centering
  \caption{Micro-Ops and data accesses savings with coarse level of work-scheduling}
  \label{tab:uop_savings}
  \resizebox{0.5\textwidth}{!}{
  \begin{tabular}{c|ccccrc}
    \hline
           &      &             & Tile Size                                             &      & uOps    & Data-accesses \\
    Layers & Type & Total $Ops$ & ${\scriptscriptstyle (\Delta O, \Delta C, \Delta N)}$ & uOps & Savings & Savings       \\
    \hline
    L2  & SCN   & 1.3e+9 & (28,16,32) & 2.5e+6 & 512x & 1.93x \\
    L12 & Conv  & 4.4e+8 & (12,16,32) & 8.6e+5 & 512x & 1.94x \\
    L24 & Dconv & 1.7e+8 & (860,8,8)  & 2.7e+6 & 64x  & 1.75x \\
    L35 & SCN   & 5.4e+9 & (212,8,16) & 4.2e+7 & 128x & 1.88x \\
    \hline
    
  \end{tabular}
}
\end{table}
We discuss challenges in adopting prior sparse-accelerators for 3D spatially sparse data processing:\newline
\bfit{A) Front-end}: for 3D spatially sparse convolution, valid input pair selection depends solely on whether the voxel is active in the receptive field in the input pointcloud and not on the weights. Further, the selection logic (Figure \ref{fig:corf_cirf_metadata}) significantly differs from the logic in any of these prior sparse accelerators. \newline%
\bfit{B) Back-End}: the back-end in these accelerators receives a list of uops, each representing a scalar multiply and accumulation operation (MAC). In 3D spatially sparse convolution, each valid input pair selection requires either a) a vector dot product (V-V) or b) a matrix-vector (M-V) multiply maximizing the data reuse when there are multiple output channels (N) in a tile. Hence, scheduling the work at the granularity of M-V would be more optimal than dispatching the work per MAC.
\bfit{Table \ref{tab:uop_savings} shows how a M-V granularity of dispatch can help to achieve a huge reduction in dispatched uops and in data accesses between compute and on-chip memory for a few select layers in Unet \cite{graham20183d}.}
We use 64KB of on-chip memory with optimal tile size and dataflow selection to minimize data accesses between on-chip and off-chip memory.

We thus present a sparse-accelerator for spatially-sparse DNNs that consists of a) a sparse scheduler based Front-end to process custom-compressed 3D metadata and b) a Back-end that dispatches matrix-vector operations which maximizes data reuse and minimizes the dispatched uops. 

\section{Hardware-Software Co-Designed Acceleration for 3D Spatial Sparsity}
\label{sec:acc_for_ss}
\subsection{Locality aware Data-structure (\textit{COIR})}
\label{sec:coir}
\begin{figure}[t]
  \centering
  \includegraphics[width=0.45\textwidth]{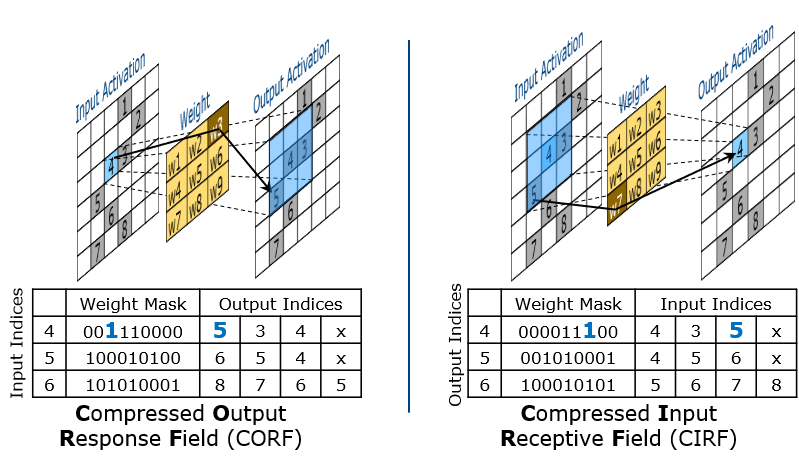}
  \caption{Two-dimensional illustration of the two flavors for \coir metadata structure for submanifold convolution operator. With 3D visual data input, output activations and weights would be in form of cuboids (instead of planes)}.
  \label{fig:corf_cirf_metadata}
\end{figure}

Workload analysis in Section \ref{sec:wl_char} provided the following insights: 1) performing convolution for all the voxels together in the receptive field (i.e. neighbourhood points) can reduce data accesses for input/output feature maps 2) storing all input point indices per output point (or vice a versa) can achieve metadata compression compared to weight wise listing of in-out pairs\cite{graham20183d}.
Metadata size savings could be significantly higher for denser pointclouds. Inspired by these insights, we propose a novel metadata structure (\bfit{COIR}) with two flavors 1) \bfit{Compressed Output Response Field (CORF)}, 2) \bfit{Compressed Input Receptive Field (CIRF)} (Figure \ref{fig:corf_cirf_metadata}).
In \textit{CORF}, each metadata entry corresponds to one unique point in input space and consists of 1) the index of the input point, 2) indices of all the points in the output space in the response field of the input point.
In addition, relative location of each neighbour is also required to select the appropriate weight index for convolution operation.
For this, a weight bit-mask is stored for each entry with '1's indicating valid neighbours and bit-locations indicating corresponding weight indices.
Similarly, each \textit{CIRF} entry contains the index of a unique output point, the indices of all the input points in the receptive field required to compute feature map for the output point and bit-masks for weight index selection.
Two flavours of metadata are motivated by the observation that
for the layers with resolution change (input and output space having different resolution), input to output mapping is not one-to-one.
\bfit{Hence, based on the sparsity in the pointcloud and the type of convolution (upscaling or downscaling) picking the right flavor could provide higher compression and data-savings over the other.}

\subsection{Point-cloud Reordering (\soarx)}
\label{sec:soar}
\textit{COIR} improves data reuse of input points with \textit{CORF} (or output points with \textit{CIRF}) by performing convolution across multiple neighbours per element access, but it doesn't ensure data reuse of neighbours across multiple input or output points.
\bfit{To maximize data reuse for neighbours as well, entries in the metadata needs to be ordered such that the entries with shared neighbours are co-located in the metadata structure and processed in close temporal vicinity.}
To achieve this, we propose \bfit{Surface Orientation Aware Reordering} (\soarx) of the pointcloud.
We first create an \textit{Adjacency Map} with the voxel index as key and a list of indices to all its neighbours as value.
When provided as input the maximum number of voxels for which data can fit in the on-chip memory, \soar divides the entire pointcloud into multiple chunks, each chunk composed of ordered voxels obeying the maximum voxels constraint.%
\begin{figure}
  \centering
  \includegraphics[width=0.50\textwidth]{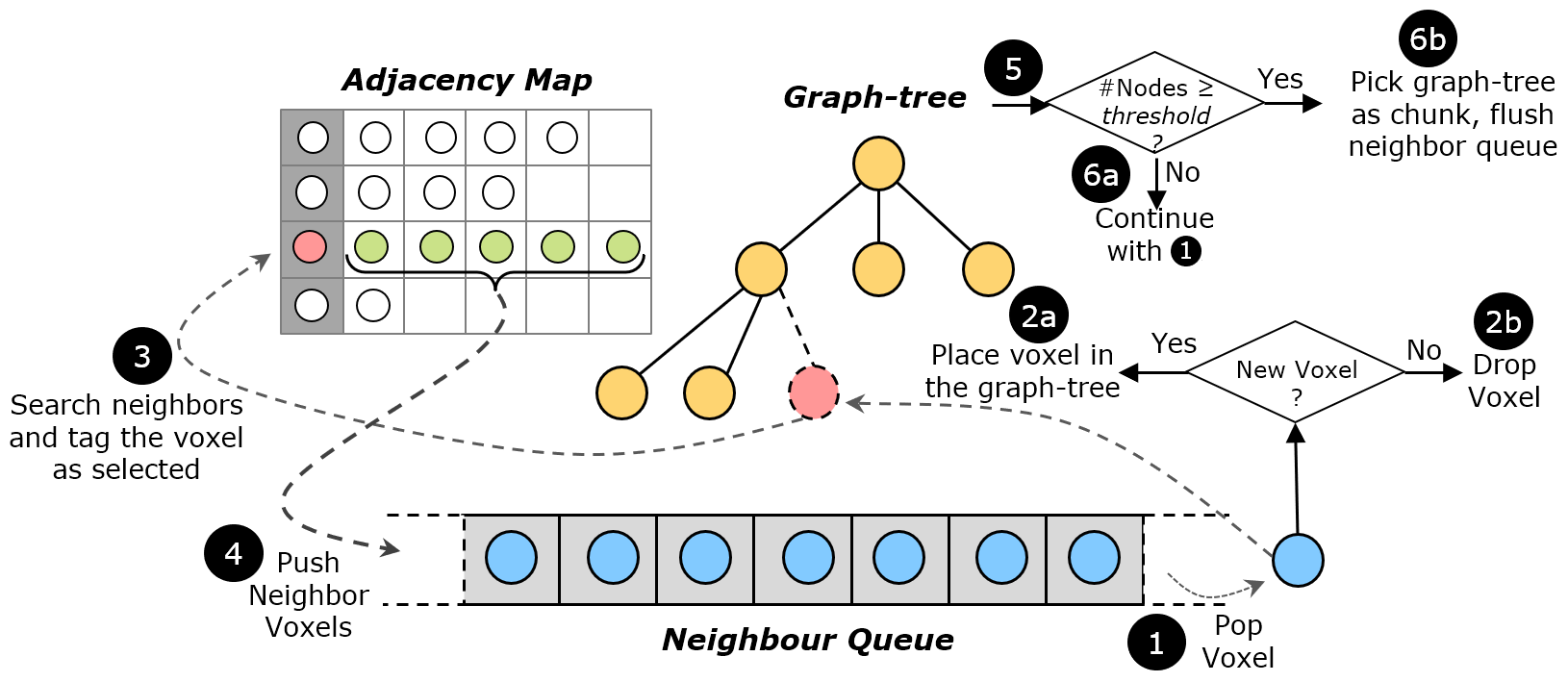}
  \caption{An illustration for dividing points in pointclouds in size constrained \textit{chunks} and reordering points in each chunk through \soarx}
  \label{fig:soar_divide}
\end{figure}

As shown the Figure \ref{fig:soar_divide}, \soar constructs an m-ary tree with each node representing a voxel and an edge connecting two neighbouring voxels such that each child node connects to only one parent even if there is more than one neighbour at parent's levels, though a parent node can connect to one or more neighbours as child nodes.
To begin, the voxel with minimum number of neighbours is selected as the root node representing a corner in the pointcloud. \soar pushes all its neighbours to a \textit{Neighbour Queue} and pops out voxels one-by-one.
If a voxel is already selected in previous chunks, it gets dropped otherwise it is inserted into the m-ary tree as a child node to the first neighbour in breadth-first order.
The inserted voxel is then tagged as selected and all its neighbours are added to the \textit{Neighbour Queue}.
This process continues till the number of voxels in the m-ary tree matches with the provided threshold, upon which the m-ary tree is selected as the desired chunk with voxels in the breadth-first order.
The root node for the next chunk is selected among voxels in the \textit{Neighbour Queue} with minimal number of neighbours and then this queue is flushed.
This process gets completed when all the voxels in the pointcloud are divided into chunks.

\subsection{Sparsity Aware Dataflow (\spade)}
\label{sec:spade_df}
Section \ref{sec:need_for_df} highlighted two key challenges of tiling 3D spatially sparse data: 1) memory requirement varies highly due to input dependent spatial sparsity 2) data accesses can not be estimated through mathematical expressions due to irregular data-structures. In this section, we propose a framework for dataflow exploration for spatially sparse data processing.

\begin{figure}
  \centering
  \includegraphics[width=0.45\textwidth]{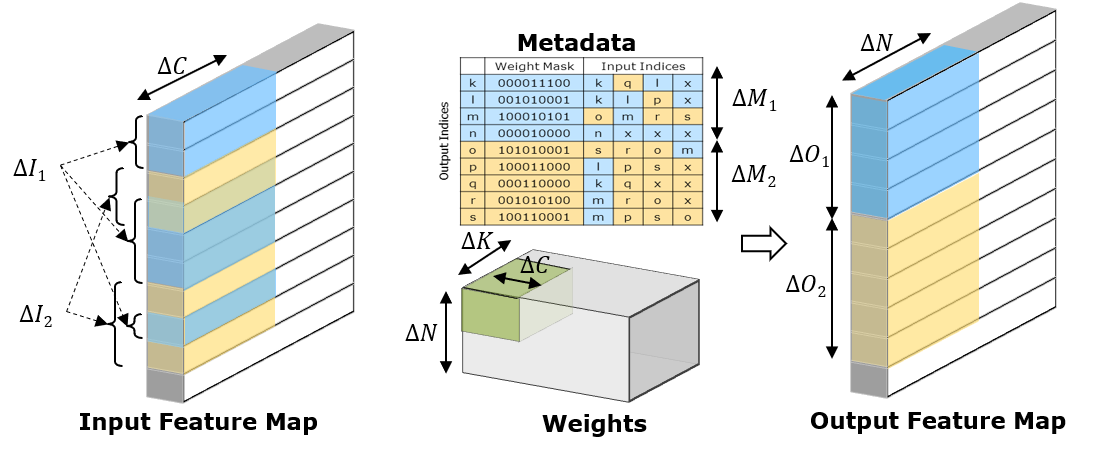}
  \caption{Tiling 1D compressed data-structure for 3D spatially sparsed data}
  \label{fig:spade_tiling}
\end{figure}

\begin{figure}[b]
  \centering
  \includegraphics[width=0.4\textwidth]{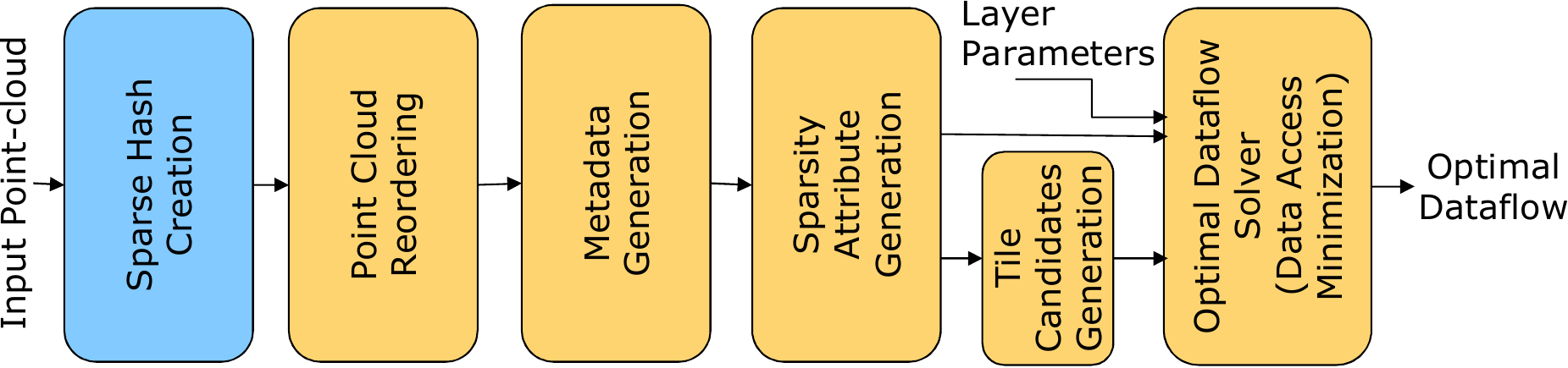}
  \caption{\spade Dataflow Exploration Framework}
  \label{fig:spade_blk_diag}
\end{figure}
For a layer $\mathcal{L}$, let the total number of input voxels, output voxels, filter size, input channels, output channels and metadata size be represented as $I$, $O$, $K$, $C$, $N$ and $M$ respectively. And if $\Delta$'s correspond to respective values in a given tile $\mathcal{T}$ (Figure \ref{fig:spade_tiling}), then the tile size $\Delta T$ can be expressed as:
\begin{equation}
  \footnotesize
  \Delta T = (\Delta I . \Delta C) + (\Delta O . \Delta N) + (\Delta K . \Delta C . \Delta N) + \Delta M\\
\end{equation}
For a number of output voxels ($\Delta O$) in the tile, required number of input voxels and metadata size could be obtained per region ($\mathcal{R}^{\Delta O}_{i}$) as:
\begin{equation}
  \footnotesize
  \label{eq:fimo}
  \begin{split}
    \Delta I = f_{I}(\mathcal{R}^{\Delta O}_{i}, \Delta O), \; \Delta M = f_{MO}(\mathcal{R}^{\Delta O}_{i}, \Delta O)
  \end{split}
\end{equation}
Tile size $\Delta T$, therefore, could be formulated as a complex function of $\Delta O$, $\Delta N$ and $\Delta C$.
There are two possible ways of tiling: 1) dynamic tiling and 2) static tiling.
In dynamic tiling, the data fetch module can keep fetching additional input feature maps as required for computing every new output feature map till the on-chip memory is full, thus each tile can have different value of $\Delta O$.
Note that to enable this, $\Delta N$ and $\Delta C$ need to be known prior to the execution and, therefore, \bfit{a joint optimization of all tiling parameters could not be performed}.
A solution to this would be to process the metadata before the DNN execution extracting out region information ($f_{I}$, $f_{MO}$) for every region and empirically choose the optimal tiling candidates by iterating over all possible values of $\Delta C$ and $\Delta N$ pairs. This would explode the dataflow exploration space and require multiple passes of metadata processing significantly degrading latency in real-time scenarios.
\begin{figure*}
  \centering
  \includegraphics[width=0.90\textwidth]{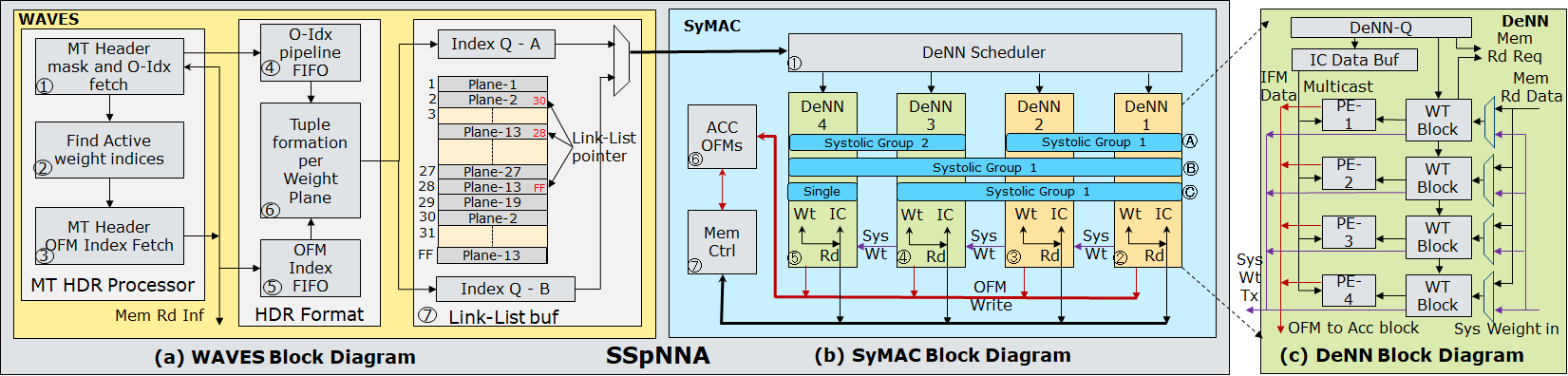}
  \caption{Micro-Architecture block diagram for \spnna, \textbf{(a)} \textbf{WAVES} group features as per weight planes, stores tuples into Link-List index queue, \textbf{(b)} \textbf{SyMAC} connects DeNNs as group of systolic arrays along the weight data, \textcircled{6} perform caching and local accumulation on hits \textbf{(c)} DeNN block where IC-data is buffered, to be reused among multiple output channels and multi casted to all the PEs.}
  \label{fig:uarch_blk_diag}
  \vspace{-0.7cm}
\end{figure*}
Thus, we \bfit{propose to decouple extraction of region information in the form of sparsity attributes and then perform dataflow exploration using these attributes via an analytical framework}.
We compute sparsity attributes, $\mathcal{SA}_{I}$ and $\mathcal{SA}_{MO}$, as function of $\Delta O$ over all regions (Eqn. \ref{eq:gimo}).
Note that $\mathcal{SA}_{MO}$ represents Average number of voxels in Receptive Field (ARF) for a given region, while $\mathcal{SA}_{I}$ takes the form of $(1 + \beta)$ where $\beta$ reflects the fraction of voxels at region boundary.
\begin{equation}
  \footnotesize
  \label{eq:gimo}
  \begin{split}
    \mathcal{SA}_{I}(\mathcal{R}^{\Delta O}_{i}, \Delta O) & = \left(f_{I}(\mathcal{R}^{\Delta O}_{i}, \Delta O)\right) \; / \; {\Delta O} \\
    \mathcal{SA}_{MO}(\mathcal{R}^{\Delta O}_{i}, \Delta O) & = \left(f_{MO}(\mathcal{R}^{\Delta O}_{i}, \Delta O)\right) \; / \; {\Delta O}
  \end{split} 
\end{equation}
We adopt static tiling with two ethods: 1) Strict Static Tiling (SST) and 2) Relaxed Static Tiling(RST). In SST, $\mathcal{SA}_{I}$ and $\mathcal{SA}_{MO}$ with the highest value across all regions are picked for each $\Delta O$ to ensure worst-case tile size allocation while under-utilizing the on-chip memory for quite a few tiles.
In RST, to improve memory utilization, we use n\textsuperscript{th} quantile of $\mathcal{SA}_{I}$ and $\mathcal{SA}_{MO}$ such that majority of the tiles fits within on-chip memory. The tiles overshooting the on-chip memory are split into two (or a next power of two) such that each sub-tile fits well into the on-chip memory.
To calculate data transfers and operations, we use $\mathcal{SA}^{Avg}_{I}$ and $\mathcal{SA}^{Avg}_{MO}$ (Eqn. \ref{eq:himo}) averaging the sparsity attributes over all the regions ($\mathbb{R}^{\Delta O}$) for each $\Delta O$.
\begin{equation}
  \footnotesize
  \label{eq:himo}
  \begin{split}
    \mathcal{SA}^{Avg}_{I}(\Delta O) & = \left(\Sigma_{\mathcal{R}^{\Delta O}_{i} \in \mathbb{R}^{\Delta O}} \mathcal{SA}_{I}(\mathcal{R}^{\Delta O}_{i}, \Delta O)\right) / \; |\mathbb{R}^{\Delta O}|, \\
    \mathcal{SA}^{Avg}_{MO}(\Delta O) & = \left(\Sigma_{\mathcal{R}^{\Delta O}_{i} \in \mathbb{R}^{\Delta O}} \mathcal{SA}_{MO}(\mathcal{R}^{\Delta O}_{i}, \Delta O)\right) / \; |\mathbb{R}^{\Delta O}|
  \end{split} 
\end{equation}
Similar to dense DNN, we choose between three types of walk-patterns ($\mathscr{WP}$) 1) Input Stationary ($IS$) 2) Output Stationary ($OS$) and 3) Weight Stationary ($WS$).
Note that Eqn. \ref{eq:fimo}, \ref{eq:gimo} and \ref{eq:himo} could also be expressed as functions of $\Delta I$ computing for $\Delta O$ and $\Delta M$ with CORF metadata structure and ARF representing as Average Response Field.
Using the analytical framework (as shown in Figure \ref{fig:spade_blk_diag}), \spade explores the entire dataflow design space to arrive upon the best tile size, walk pattern and metadata structure ($\mathscr{MD}$) for each layer in the network given an input pointcloud.
\spade's Analytical framework minimizes data accesses $\mathscr{DA}$ (Eqn. \ref{eq:data_acc}) between the on-chip memory and the off-chip memory over the entire dataflow design space $\mathcal{D}=\{(\mathcal{T},\mathscr{WP},\mathscr{MD})\}$.
\begin{equation}
  \footnotesize
  \label{eq:data_acc}
  \begin{split}
    \mathscr{DA}&\left(\mathcal{D}\right) = F_{WS}\left(\mathscr{WP},\ceil{O/\Delta O}\right).\left(C.N.K\right)\\
    & + F_{IS}\left(\mathscr{WP},\ceil{N/\Delta N}\right).\left(\mathcal{SA}^{Avg}_{I}(\Delta O).O.C\right) \\
    & + F_{OS}\left(\mathscr{WP},\ceil{C/\Delta C}\right).\left(O.N + \mathcal{SA}^{Avg}_{MO}(\Delta O).O\right) \\
    \text{where, }&\;\;F_{X}(Y,Z) = {1\;if\;{\left(Y = X\right)},\;else\;{Z}}
  \end{split}
\end{equation}

\subsection{3D-Sparse-NN-Core Micro-architecture: \spnna}
\label{sec:uarch}
Section \ref{sec:need_for_acc} described the distinctive requisites of a custom accelerator for 3D spatial sparsity, especially contrasted with prior sparse accelerators. We describe in this section, the microarchitecture of the \spnna (\textbf{S}patially \textbf{SP}arse \textbf{N}eural \textbf{N}etwork \textbf{A}ccelerator) core which lies at the heart of our AccSS3D solution.
The primary requisite for the hardware accelerator (HWA) is to process a tile (Figure \ref{fig:spade_tiling}), convert from sparse to dense representation by restructuring the tile data in the Front-end and perform dense compute in the Back-end.

\textbf{Top-level Overview:}
The \spnna HWA is comprised of two major blocks (a) \textbf{WAVES Front-end}- \textbf{W}eight plane based \textbf{A}ctive \textbf{V}oxel \textbf{E}xecution \textbf{S}cheduler (b) \textbf{SyMAC Back-end}- \textbf{S}ystolic and \textbf{M}ulticast based \textbf{MAC} \textbf{C}omputation in Figure \ref{fig:uarch_blk_diag}. Metadata (MT) is partitioned into header and feature data, where header stores \coir bit-masks and feature data stores the feature indices and its data. HWA also has a memory arbiter, a configuration and a control block. The Global event controller initiates execution after loading the L1 and configuring the HWA. Upon start, the WAVES scheduler formats the metadata and then triggers SyMAC for channel-wise computation and output element accumulation, during which, WAVES starts working on formatting the next set of data.

The \textbf{WAVES Front-end} rearranges the spatially distributed voxels along weight plane (Figure \ref{fig:uarch_blk_diag}-a). Its subblock MT HDR Processor fetches weight mask and corresponding IFM and OFM indices. By using smart-lookup, it finds weight index for 4 active neighbor voxels per cycle. The HDR Format block forms tuples by grouping 4 features per weight plane, using 27 blocks to manage all weight planes together. Link-List based buffer is used to provide dynamic memory space allocation for weight planes, enabling more storage for weights with more active voxels.
This design choice was motivated by the observation that fixed allocation of resources per weight plane leads to significant under-utilization, since it depends on \textit{ARF}. By allocating 1 FIFO per weight plane for tuple storage, higher occupancy for all the FIFOs cannot be achieved. This under-utilization can be visualized as a wavy line touching the top element in each FIFO. %
The Link-List design helps increase the utilization by dynamic allocation of more resources to planes with higher number of active neighboring voxels, \bfit{allowing us to accommodate 1.5X-2X more metadata lines} in the same size of memory internal to the \spnna.

The \textbf{SyMAC Back-end} increases the data reuse within the HWA by connecting multiple compute blocks systolically, multicasting the input features to a set of PEs and accumulating data within the PEs as per channel length. Figure \ref{fig:uarch_blk_diag}(b) shows three options for systolic groups as \textcircled{A}, \textcircled{B}, \textcircled{C}, which can be dynamically selected for weight plane grouping in the WAVES. Figure \ref{fig:uarch_blk_diag}(b)\textcircled{6} ACC OFM block has local buffering with cache lookup capability which helps in reducing memory transactions for elements with cache hit.
Figure \ref{fig:uarch_blk_diag}(c) shows a DeNN block where input features are buffered and reused for all output channels, by convolving with different weights. PEs are implemented using tree structure where they can perform dot-product on IEEE754 Full Floating-Point numbers and accumulate locally along the input channels.%

4-DeNN configuration with 4 PEs per DeNN computes 4 elements per PE per cycle allowing \spnna to support 64-MUL operations per cycle. Changing \spnna configuration to 8 DeNNs, working in two systolic groups of 4-DeNN each, doubles performance to 128 MUL operations per cycle, but does not require any more memory ports for weights.
\textit{\textbf{Through conscious design choices we reduce bandwidth requirement by 68\%}}: a) Local input buffer sharing and Systolic weight connection reduce bandwidth requirement by 37\% and 25\% respectively, b) local accumulation within PEs for $\Delta C$ and $\Delta N$ greater than 15 provides another 6\% reduction.

\begin{figure} 
	\centering
	\includegraphics[width=0.48\textwidth]{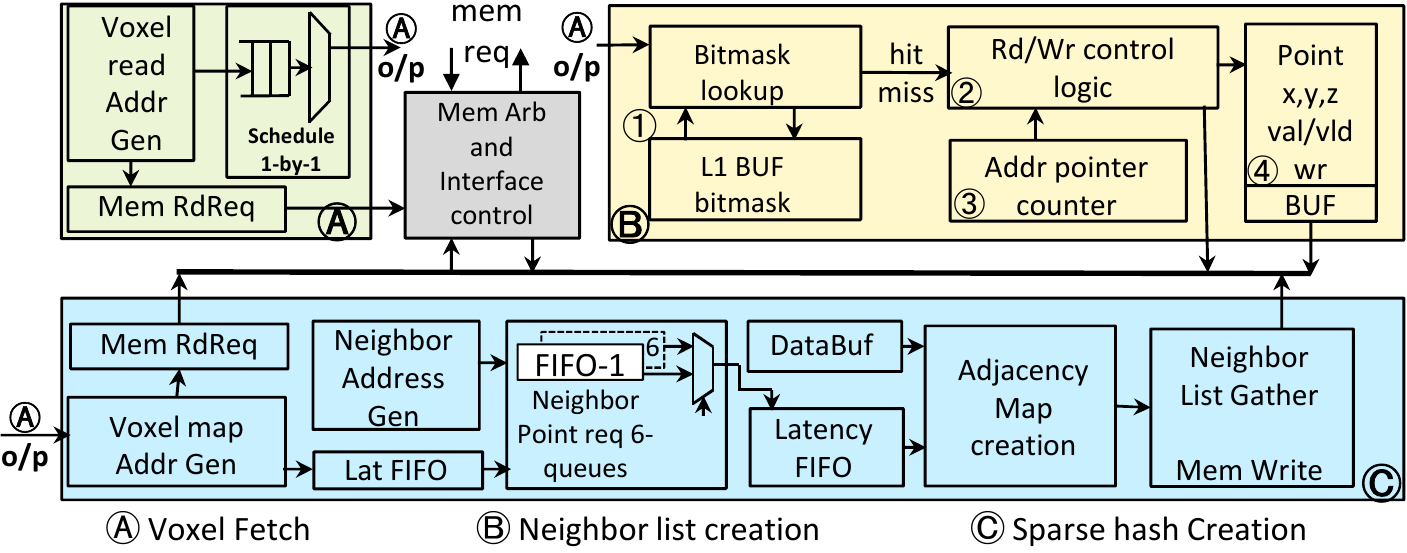}
	\caption{Micro-Architecture block diagram for \admac, \textcircled{A} is common block streaming voxels from memory, \textcircled{B} prepares adjacency map to help \textcircled{C} create Metadata with minimal memory reads}
	\label{fig:adj_map_accl}
\end{figure}
\subsection{Adjacency Map accelerator Micro-Architecture - \admac}
Section \ref{sec:soar} described the requirements for reordering pointcloud and Section \ref{sec:spade_df}, Figure \ref{fig:spade_tiling} described metadata structure for tiling. We describe in this section, the Micro-Architecture of \textbf{\admac}-\textbf{A}djacency map and \textbf{M}etadata \textbf{Ac}celerator core in Figure \ref{fig:adj_map_accl}. Input voxels (x,y,z) are fetched serially by block \textcircled{A} in Figure \ref{fig:adj_map_accl}. Block \textcircled{B} creates lookup table for all the active voxels. For faster lookup, \admac maintains hierarchical lookup table, at level one it encodes active voxels at higher granularity (called voxel 3D groups) and at second level it stores active information and corresponding memory address per voxel. Voxel information is stored in an 8-banked memory where \texttt{bankID} is encoded using \{y[2], z[1:0]\}. Within the bank, voxels are hashed so that each memory read of 64 Bytes can provide information for 16 voxels as per \{y[1:0], x[1:0]\} addressing. \bfit{This specific hashing helps in reading 26 neighboring voxels in a single cycle}, with the exception of boundary voxels. Block \textcircled{C} creates Adjacency List for the voxels in the memory using the Sparse hash from \textcircled{B}. Blocks \textcircled{C}\textcircled{1},\textcircled{2},\textcircled{3} compute address of neighbors, reads voxel information and write to memory after packing data as per metadata structure.

\section{Scale-up Architecture for 3D Spatial Sparsity}
\label{sec:arch}
\begin{figure}
	\centering
	\includegraphics[width=0.45\textwidth]{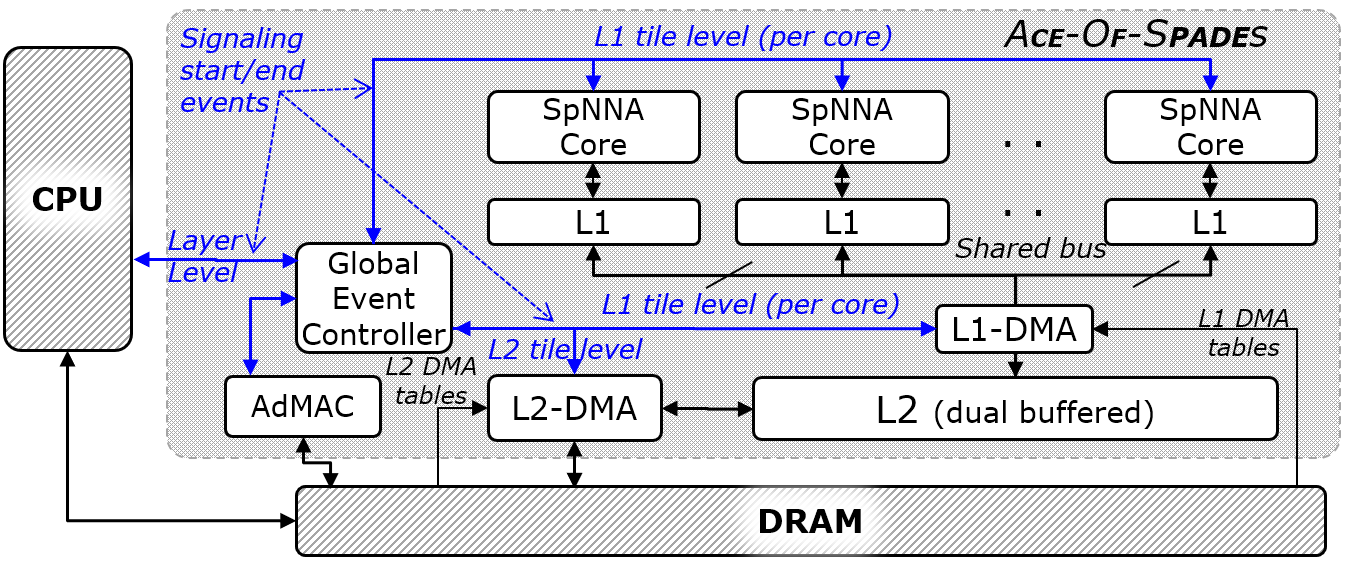}
	\caption{Proposed architecture for spatial sparsity: \aos}
	\label{fig:arch_block_diagram}
\end{figure}

In this section, we present \aos, an architecture (Figure \ref{fig:arch_block_diagram}) targeting 3D spatially sparse DNN applications. 

\subsection{Overall Architecture}
\subsubsection{\textbf{On-die memory architecture}}
Multi-level memory hierarchies have been widely adopted as they allow high bandwidth short-distance reuse at smaller inner level memories with lower latencies, while longer-distance reuse are captured at higher capacity by outer memory levels. 
The two levels of memories in the AccSS3D architecture are managed as scratchpads by the software to orchestrate the 3D sparse-CNN execution following the optimal directives of the \spade.
A scratchpad-based architecture with multiple levels of memory, requires compute and data-transfer needs to be effectively synchronized to maximize compute and bandwidth utilization.
As a result of the locality-aware tile-based execution as described in section \ref{sec:spade_df}, there is significantly higher space sensitivity at L1 compared to L2.
This is because when processing similar number of voxels, the unique-OFM to unique-IFM ratio is typically highly skewed (away from 1) for L1 when compared to L2.
This results in higher sensitivity for reuse opportunity to memory size at L1 as compared to L2.
This precludes double-buffering at L1.
\textit{Hence, we choose distinct compute and data-exchange phases as in \cite{graphcore}}.
All data transfers from between the shared-L2 and a \spnna core's L1 are blocked when the core is active, and the \spnna core is idled when data is being exchanged between its L1 and the shared L2.

\subsubsection{\textbf{Scaled-up multi-sparse-NN Core architecture}}
To compensate for the increased latency due to sequential phases, we adopt an overlapped tile execution model across the multiple \spnna cores in our scaled-up architecture (Figure \ref{fig:async_exec_load_balance}-a).
Using a shared bus at the L1-L2 interface, tile data is transferred between the shared-L2 and a \spnna core's L1 memory while other cores are in their compute phases and this continues in round-robin mode as each core enters its data exchange phase.

\subsubsection{\textbf{Data transfers between L1, L2 and DRAM}}
\label{sec:data_tx_l1_l2}
We employ a DMA-based architecture to accomplish data transfers in the data-exchange phases.
DMAs are triggered through a global hardware-based event controller, with the software specifying data-movement details such as source/destination addresses and number of bytes through the DMA engines tables.
We provision two DMA engines: one each for the L1-L2 and L2-DRAM interfaces respectively.
Based on the selected metadata structure (\textit{CORF} or \textit{CIRF}) - either one of the datatypes between IFM and OFM is accessed in order, while the other can be un-ordered (see Figure \ref{fig:spade_tiling}).
We use block transfers programmed as a single DMA table entry for the entire tile for the datatype with ordered accesses. For the other datatype, we use DMA table entries per voxel level. Since weights are dense, a block transfer is sufficient. DMA transfers for all tiles of a layer are chained in a pre-selected order as this allows limiting CPU/software intervention to only once per layer.

\subsubsection{\textbf{Load Balancing with Multi-core}}
\begin{figure}
	\centering
	\includegraphics[width=0.5\textwidth]{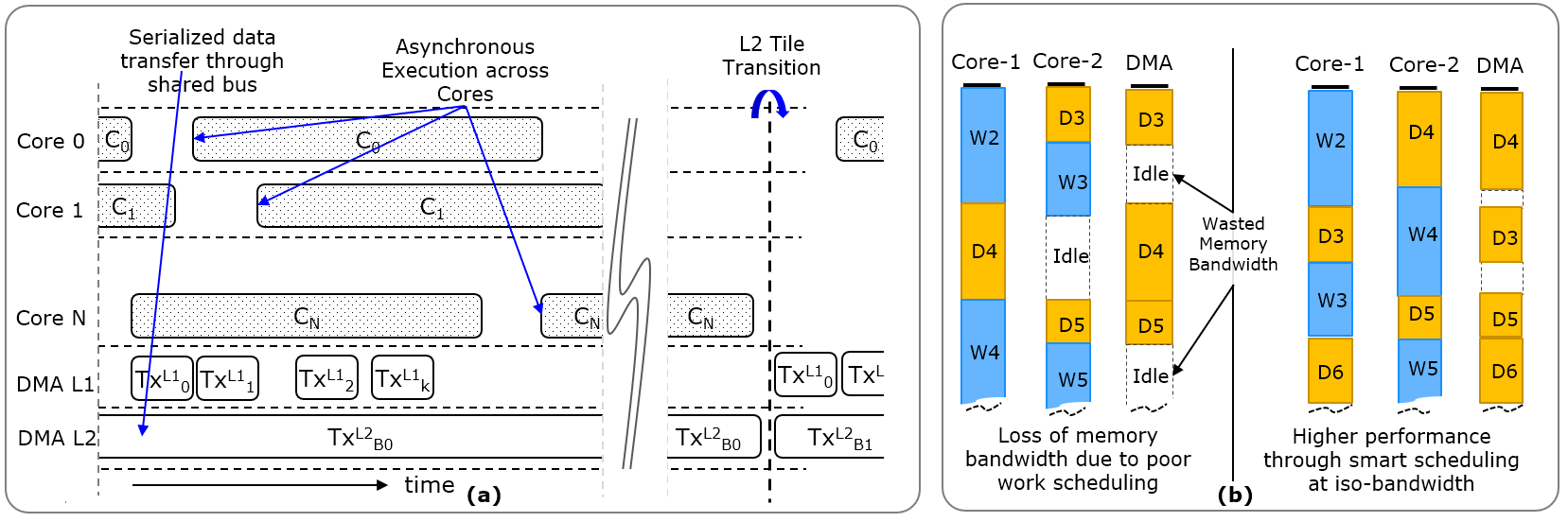}
	\caption{(a)Execution flow orchestrating asynchronous scheduling of sparse-NN cores with serialized DMA transfers. (b) An illustration of improved performance with smart work scheduling. W$_x$ and D$_x$ represents processing-work and data-transfer respectively as associated with the tile $\mathcal{T}_{x}$.}
	\label{fig:async_exec_load_balance}
\end{figure}

Though \spade ensures uniform tiling (equal $\Delta N$, $\Delta C$ and $\Delta O$ or $\Delta I$), due to region dependent sparsity (varying $\mathscr{SA}_{MO}$), operations per tile could differ.
Asymmetric core execution times and data transfer period can introduce idle periods for both processing cores and the DMA engine (Figure \ref{fig:async_exec_load_balance}b-left).
To address this, we sort all the spatial tiles based on 'Ops-per-tile' in descending order.
Based on parallelism along $N$, cores are grouped with each group to process different $N$-tiles.
The sorted spatial tiles are scheduled in order and round-robin fashion on cores in each group maximizing concurrent execution (Figure \ref{fig:async_exec_load_balance}-b-right).

\subsection{Dataflow for Multi-level memory hierarchy}
\label{sec:df_mult_mem}
Multi-level memory hierarchies pose significant challenges for dataflow optimization, as minimization of data-transfers is required at every level of the memory hierarchy to enable best performance at lowest hardware and energy cost.
Inspired by dense DNN dataflow optimizers \cite{chen2018eyeriss, yang2016systematic, hegde2018morph}, we adopt hierarchical dataflow search starting from outermost to innermost level of the on-chip memory. One of the drawbacks of hierarchical search is the likelihood of picking a globally suboptimal dataflow.
For example data access minimization at $L_q{\leftrightarrow}L_{q+1}$ memory interface may choose tile candidates with lower spatial data reuse and higher temporal data reuse, which could increase data accesses at inner interface $L_{q}{\leftrightarrow}L_{q-1}$.
To address this, we propose \bfit{Constrained Access based Reuse Opportunity Maximization (CAROM)} to maximize spatial reuse while keeping the number of data accesses at $L_{q}{\leftrightarrow}L_{q+1}$ interface lower than a maximum threshold value ($\mathscr{DA}_{th}^{L_{q}}$) without being bandwidth constrained.
For this, \textit{CAROM} first identifies a set of dataflow candidates $\mathbb{D}^{L_{q}}$, such that:
\begin{equation}
  \small
  \begin{split}
    \mathbb{D}^{L_{q}} = & \left\{\mathcal{D}_{i}: \mathscr{DA}_{\mathcal{D}_i}^{L_{q}} \le \mathscr{DA}_{th}^{L_{q}}\right\} \cup \left\{{argmin}_{\mathcal{D}_{i}}(\mathscr{DA}_{\mathcal{D}_i}^{L_{q}})\right\}\\
  \end{split}
\end{equation}
where, $\mathscr{DA}_{th}^{L_{q}}$ is computed based on number of operations to be performed on the working set at $L_{q+1}$, total compute (Ops/sec) available to an instance of memory level $L_{q}$ and bandwidth at $L_{q}{\leftrightarrow}L_{q+1}$ interface:
\begin{equation}
  \small
  \begin{split}
    \mathscr{DA}_{th}^{L_{q}} =  (Ops^{L_{q}} \times {BW}^{L_{q}}) / Total\_Comp^{L_{q}}\\
  \end{split}
\end{equation}
\begin{equation}
  \small
  \begin{split}
    \text{where,} \;Ops^{L_{q}} = \mathscr{SA}_{MO}^{Avg}\left(O^{L_q}\right).O^{L_q}.N^{L_q}.C^{L_q}
  \end{split}
\end{equation}
\textit{CAROM} then picks the optimal dataflow over the set $\mathbb{D}^{L_{q}}$ maximizing reuse opportunity ($\mathscr{RO}^{L_{q-1}}$) for $L_{q-1}$ level. 
\begin{equation}
  \small
  \begin{split}
    \mathcal{D}_{opt}^{L_{q}} = {argmax}_{\mathcal{D}_{i} \in \mathbb{D}^{L_{q}}}\left(\mathscr{RO}_{i}^{L_{q-1}}\right)
  \end{split}
\end{equation}
Since for a given working set, reuse opportunity is proportional to the total number of operations to be performed on the working set, $Ops^{L_{q-1}}$ is used for reuse opportunity maximization.
The optimal tile candidate at a memory level $L_{q}$ acts as the working set for level $L_{q-1}$.
And, therefore, \bfit{CAROM continues to pick optimal dataflow(s) from outer levels to inner levels except the innermost level as per the above criterion.
For the innermost memory level, it selects the optimal dataflow by minimizing data accesses} (${{argmin}_{\mathcal{D}_{i}}\left(\mathscr{DA}_{i}\right)}$).

To improve data locality across all the memory levels through pointcloud reordering, \soar (Section \ref{sec:soar}) is extended to perform hierarchical pointcloud ordering starting from innermost to outermost levels.
Given optimal tiling from \spadex, \soar groups the entire pointcloud in chunks and finds optimal ordering of points in each chunk based on tilling parameters for the innermost memory. Reinterpreting each chunk as a point, \soar is reapplied to recursively group these chunks into super-chunks with optimal ordering based on tile parameters of the outer memory level, till the outermost level.

\subsection{Minimizing \spade latency overhead}
\label{sec:metasparsity}
\begin{figure}
	\centering
	\includegraphics[width=0.45\textwidth]{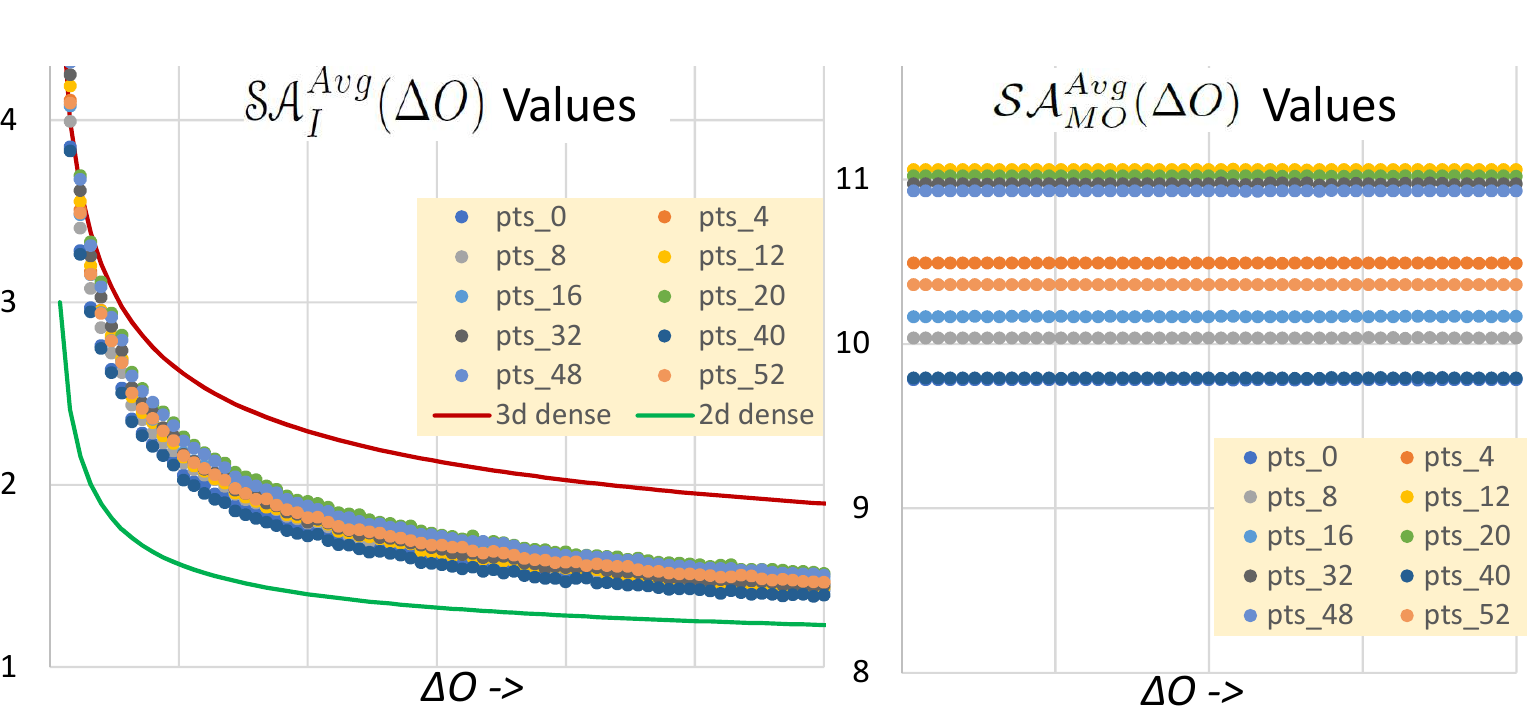}
	\caption{Sparsity attributes for different points clouds}
	\label{fig:msa_illustration}
\end{figure}

\begin{figure}
  \centering
  \includegraphics[width=0.45\textwidth]{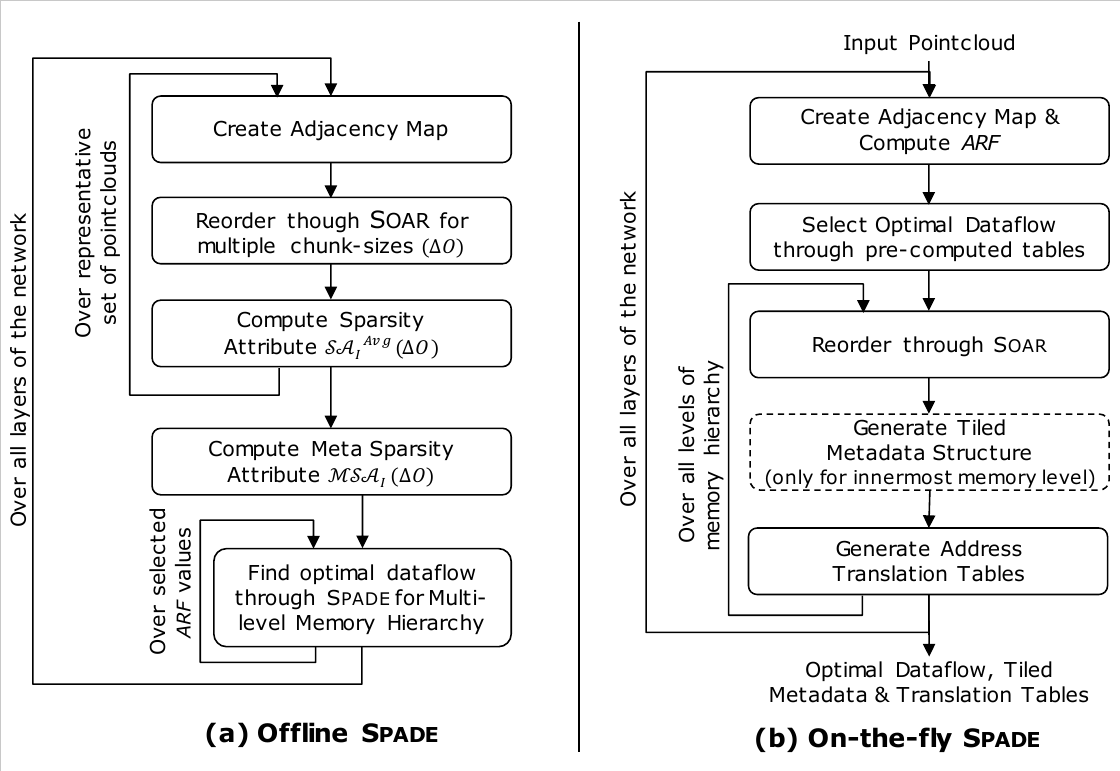}
  \caption{Processing flow for offline mode and on-the-fly mode of \spadex}
  \label{fig:spade_proc_flow}
\end{figure}

As described in sections \ref{sec:spade_df} and \ref{sec:df_mult_mem}, \spade requires pre-processing of input pointcloud data to extract sparsity attributes ($\mathscr{SA}$'s) to perform dataflow exploration. Extraction of $\mathscr{SA}$'s
could add significant overhead to end-to-end latency.
To minimize this, we explore: 1) if the sparsity attributes could be categorized into two sets: \textit{a)} common attributes which are consistent across pointclouds - referred to as \bfit{Meta Sparsity Attributes ($\mathscr{MSA}$)},  \textit{b)} \bfit{Input Specific Attributes ($\mathscr{ISA}$)} which varies highly across pointclouds; and, 2) whether by using $\mathscr{MSA}$, optimal dataflow candidates could be pre-computed for selected binned values of $\mathscr{ISA}$.
Thus, we correlate $\mathscr{SA}^{Avg}$ over randomly picked pointclouds (Figure \ref{fig:msa_illustration}) and observe that: \newline
1. $\mathscr{SA}_{I}^{Avg}(\Delta O)$ exhibits high variance across various values of $\Delta O$, but follows similar pattern across pointclouds. Also, $\mathscr{SA}_{I}^{Avg}(v)$ shows a high correlation with $(\alpha_m/\sqrt[\leftroot{-2}\uproot{2}m]{v})$ representing surface area to volume ratio of $m$-dimensional cube with volume $v$ and faces $\alpha_m$. \newline
2. $\mathscr{SA}_{MO}^{Avg}(\Delta O)$ which represents the \textit{ARF}, remains constant with $\Delta O$ for a pointcloud, but varies across pointclouds.\newline
Based on above observations, we propose an offline version of \spade to use $\mathscr{MSA}_{I}$ as the meta sparsity attribute and \textit{ARF} as $\mathscr{ISA}$. The $\mathscr{MSA}_{I}$ is computed over a representative set of pointclouds $\mathbb{P}$ using Eq. \ref{eq:msa_i}. We generate tables of optimal dataflow candidates with \textit{ARF} as table index, for all network layers. For tile allocation, we use $90$-quantile of $\mathscr{SA}_{I}^{Avg}(\Delta O)$ along with RST as described in \ref{sec:spade_df}. 
Note that \bfit{the proposed semi-offline mode of dataflow exploration strikes a conscious balance between DNN execution performance and runtime latency overhead of sparse dataflow selection}.
\begin{equation}
\small
\label{eq:msa_i}
\mathscr{MSA}_I^{Avg}(\Delta O) = \left(\Sigma_{p\in\mathbb{P}}{\mathscr{SA}_{I,p}^{Avg}}(\Delta O)\right) \; / \; |{\mathbb{P}}|
\end{equation}
To minimize latency overhead, \spade is deployed as two components (Figure \ref{fig:spade_proc_flow}): 1) \bfit{offline-\spade} - to generate a table of optimal dataflows for multiple selected \textit{ARF} values,  and 2) \bfit{on-the-fly \spade} (OTF-\spade) - to reorder a given input pointcloud and select optimal dataflow for the input.
Though \textit{Adjacency Map} and \textit{COIR} metadata contain a similar data structure, due to tiling and reordering, entries in \textit{Adjacency Map} are re-grouped to create tiled metadata ($\Delta M$) based on tiling parameters of innermost memory level.
To transfer data between memory levels, required DMA tables (as mentioned in section \ref{sec:data_tx_l1_l2}) are generated prior to DNN execution.
To further hide the latency for OTF-\spade, DNN execution is kicked-off just after the OTF-\spade completes processing for the first layer. Since OTF-\spade processing for the rest of the layers does not depend on DNN execution of previous layers, both the threads, OTF-\spade and DNN execution, proceed independently without further need for synchronization.

\section{Evaluation}

\subsection{Setup, Methodology and Workloads}
We evaluate AccSS3D using a whole chip performance and energy model. We design, synthesize \spnna core with 64KB L1 memory using Synopsys DC \cite{synopsysdc} at 1GHz clock, perform Place and route using Synopsys IC Compiler II \cite{synopsysic}. L2 memory of 2$\times$1MB is constructed hierarchically, using sub-arrays of size 16KB each, with 4 banks per instance, 4 sub-banks per bank and 4 sub-arrays per sub-bank.
Power consumption for compute and on-chip buffers is estimated using Synopsys PrimeTimePX from SystemVerilog (SV) simulation.
Interconnect energy is computed using estimated wire lengths and added to L1 and L2 access energy estimates.
Area and energy cost for \admac is estimated using micro-architecture accurate model and counting data access events.
DRAM power is taken from Micron power calculator \cite{micron2020power} for DDR4-2660.\newline
\bfit{Performance}: To measure AccSS3D performance, we obtain execution cycles for \spnna core for each tile processing in every network layer through SystemVerilog simulation. For end-to-end latency estimation, we simulate multi-core asynchronous execution model through a detailed analytical framework feeding per tile execution time from the SV simulation and including processing time for OTF-\spade (Section \ref{sec:metasparsity}).
CPU (Intel-i7-8700K @3.7 GHz) performance is measured through VTune \cite{amplifier2019intel} and power is estimated using PSST tool \cite{mubeen2018workload}. GPU (Nvidia GeForce-GTX-1080 @1.6GHz) performance is obtained form visual profiler \cite{nvidia2016nvidia} and power is reported from \textit{nvidia-smi}. We denote single core CPU, 4-core CPU and GPU software baseline performance as 1-CPU, 4-CPU and GPU. \newline
\bfit{Workloads and Datasets}: \textit{We evaluate AccSS3D on three applications of 3D visual analytics 1) 3D semantic segmentation 2) 3D object detection 3) 3D scene completion picking three state-of-the-art workloads 1) SCN \cite{graham20183d}, 2) PV-RCNN \cite{shi2019pv} and 3) SGNN \cite{dai2019sg} one from each application. We choose 3D pointclouds from ScanNet \cite{dai2017scannet} and Waymo Open Dataset \cite{sun2019scalability} (Table \ref{tab:dataset_size}, rows 5 and 6) for indoor and outdoor scenarios.}

\subsection{Scaled-up Configuration for AccSS3D}
\begin{figure}
  \centering
  \includegraphics[width=0.40\textwidth]{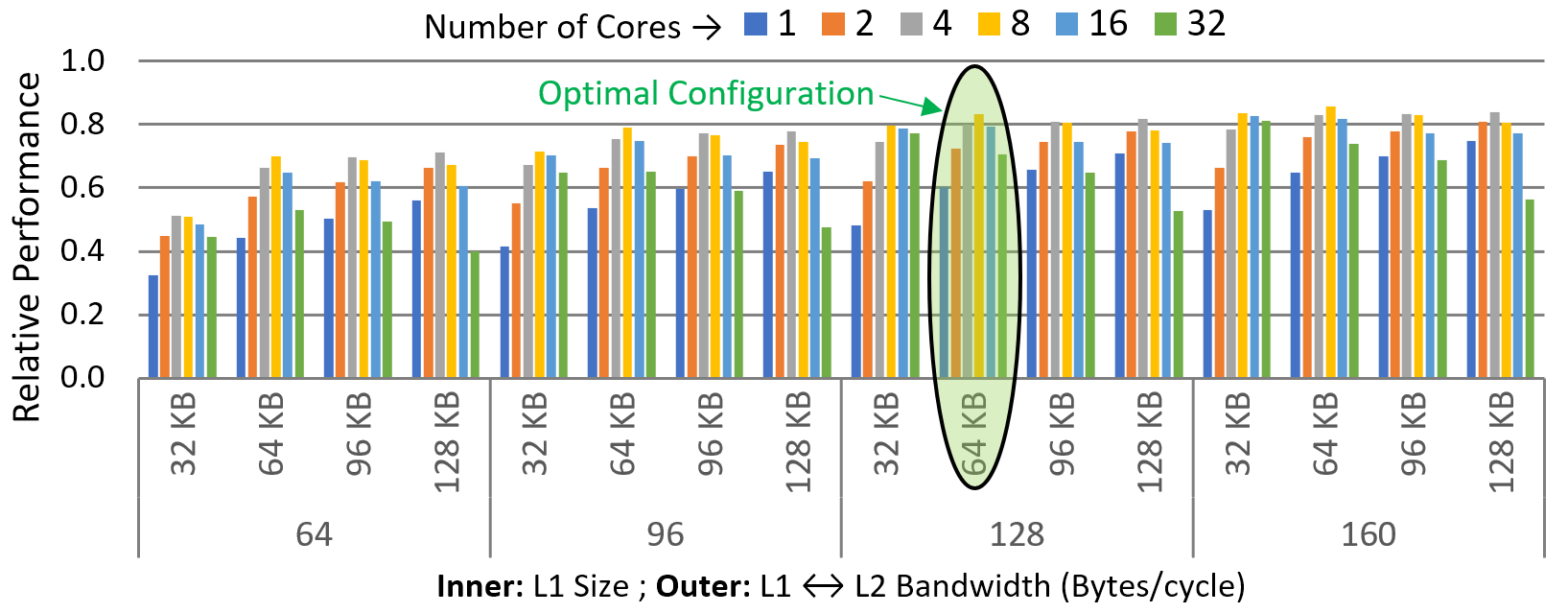}
  \caption{Performance Sensitivity to L1 memory size, L1$\leftrightarrow$L2 bandwidth and core count at an L2 size of 2 MB and DRAM bandwidth of 48 Bytes/clk.}
  \label{fig:l1_core_sel_opt_arch}
\end{figure}

\begin{figure}
  \centering
  \includegraphics[width=0.40\textwidth]{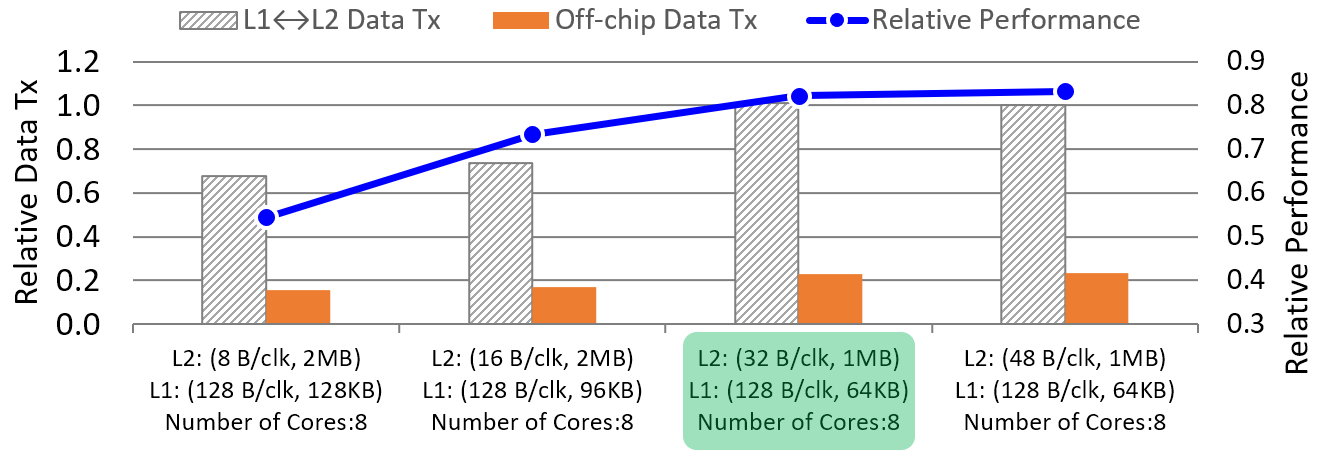}
  \caption{Relative performance, on-chip and DRAM data transfers with optimal architecture configuration for select DRAM memory bandwidth points}
  \label{fig:opt_arch_sel_off_chip_bw}
\end{figure}

We describe a 1024 MACs based AccSS3D architecture that achieves a 50x execution-time speed-up at an operating frequency of 1 GHz compared to 1-CPU.
Given the prohibitively large architecture configuration space, we adopt a hierarchical framework to arrive at the most optimal configuration.
Firstly, the optimal L2 memory size is selected such that total DRAM accesses do not exceed $1.5\times$ of the total data footprint accumulated across layers while avoiding bandwidth bottlenecks at DRAM interface for a given DRAM bandwidth.
A joint optimization is then performed to optimize L1 memory size, L1$\leftrightarrow$L2 bandwidth and number of the \spnna cores. Figure  \ref{fig:l1_core_sel_opt_arch} shows performance sensitivity to these parameters for a chosen pair of L2 memory size and DRAM bandwidth.
Performance scales with cores up to a certain number owing to a reduction in idle time during data-transfer phase.
With further increase in cores, data replication for shared datatype dominates, degrading the performance. At higher L1$\leftrightarrow$L2 bandwidth and larger L1 size, data-transfer time reduces significantly favoring lower core count configuration.
Similarly, we obtain the optimal architecture configuration for each DRAM bandwidth point (Figure \ref{fig:opt_arch_sel_off_chip_bw}) and select the optimal bandwidth that maximizes performance.
Since the L2 is dual-buffered, we provision for 2x the required size.
Figure \ref{fig:res_area_analysis}-Right lists the optimized architecture parameters.
\begin{figure*}
  \centering
  \includegraphics[trim=27 27 25 20, clip, width=1.0\textwidth]{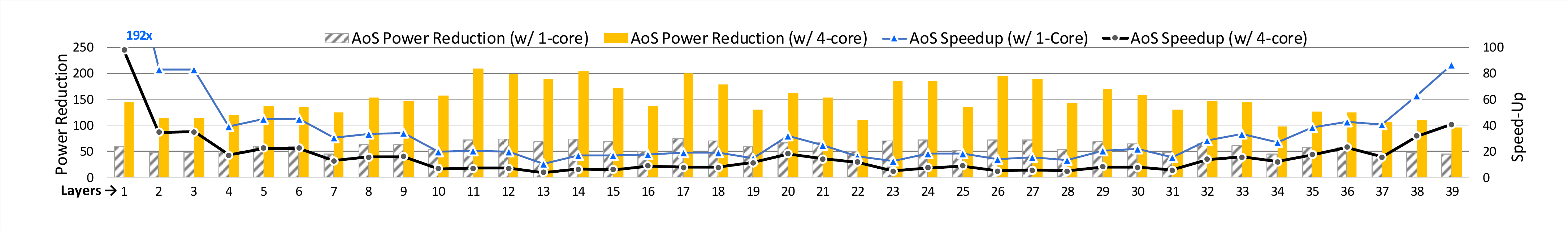}
  \caption{\textbf{Layer-wise Power Reduction and Performance} for SCN/ScanNet with AccSS3D for 3D sparse convolution over 1-CPU and 4-CPU \cite{graham20183d} (setting OpenMP threads to 1 and 4). Hyper-threading was disabled during measurement. Idle power was subtracted from total power to get workload power.}
  \label{fig:res_power_speed_up_energy}
  \vspace{-0.75cm}
\end{figure*}

\begin{table}
  \centering
  \caption{Speed-up \& Energy Savings with AccSS3D on SCN/ScanNet} %
  \label{tab:scn_savings}
  \begin{tabular}{c|l|l|l|l}
    \hline
    \multirow{2}{*}{} & 
      \multicolumn{2}{c|}{DNN Only} &
      \multicolumn{2}{c}{End-to-end} \\
      \cline{2-5}  & 1-CPU & 4-CPU & 1-CPU & 4-CPU \\
    \hline%
    Speed-up &   36.6x &   16.8x  & 23.7x  & 11.8x  \\
    Energy Savings  & 2079.0x & 2232.0x  & 23.2x  & 24.8x  \\
    \hline
  \end{tabular}
\end{table}

\subsection{Power, Performance and Area Analysis}
\begin{figure}
	\centering
	\includegraphics[width=0.42\textwidth]{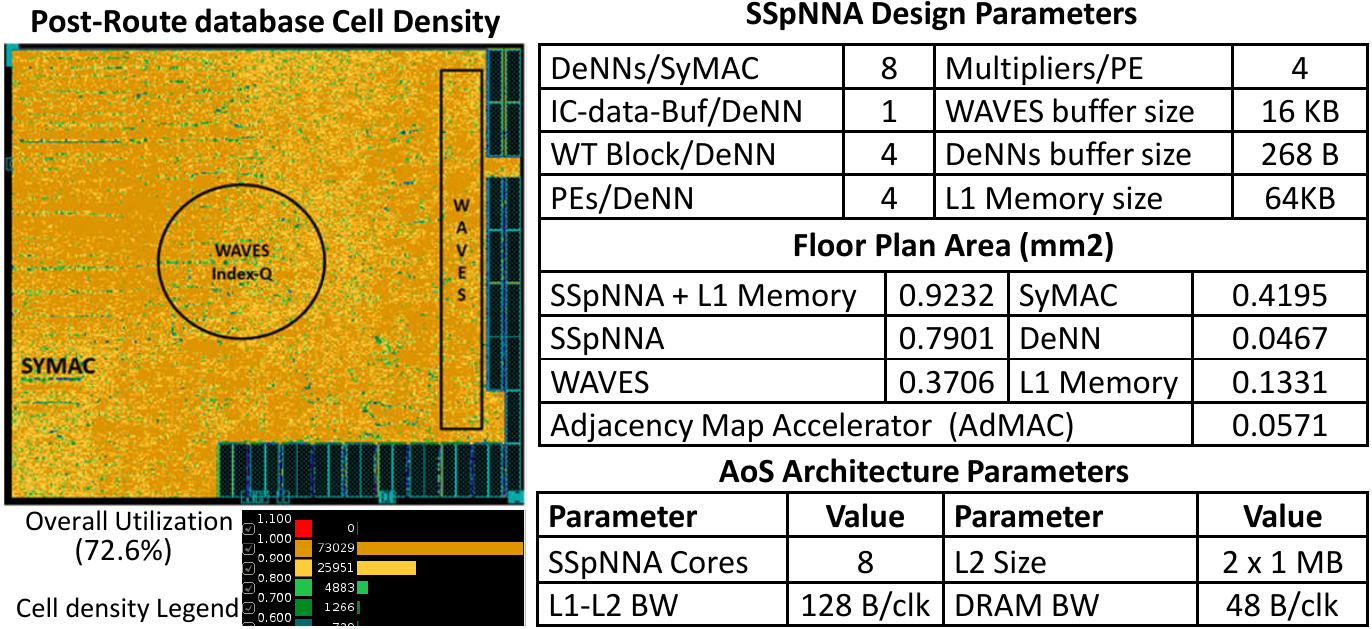}
	\caption{Left: Physical placement and area utilization of \spnna, Right-Top: 16nm \spnna Design parameters, Right-Bottom: Architecture Parameters.}
	\label{fig:res_area_analysis}
\end{figure}

As described in Section \ref{sec:uarch}, the \spnna core can be scaled with the number of internal DeNN instances. %
The design parameters, area breakup for the major blocks on a typical 16nm process and the local buffering details in the \spnna core for both configurations are shown in Figure \ref{fig:res_area_analysis}. With dual 8KB of local buffer we were able to hide the execution latency of WAVES MT formatting. Figure \ref{fig:res_area_analysis} also shows the physical placement for the \spnna core where we achieved a high utilization 72.6\% by placing the logic blocks as dictated by the internal dataflow. The energy consumption of the \spnna compute core and other local storage contributes to $\sim$50\% of total energy and remaining 50\% is attributed to SRAM accesses. 70\% of the logic power is consumed by the clock network whereas sequential and combinational cells consume 5\% and 25\% respectively.
For \textit{\admac}, energy is dominated by DRAM reads. Local buffer access and logic energy contributes to only 2\% of total energy.

Figure \ref{fig:res_power_speed_up_energy} shows layer-wise speed-up and power reduction for the 3D sparse convolution operation in the SCN network with ScanNet.
Speedup for initial and last few layers reaches up to \textbf{80x} over 1-CPU.
As \spade adopts spatial sparsity aware tiled execution, data accesses are reduced significantly.
With tiled metadata, ordered data transfers through DMAs and asynchronous execution model, data transfer latencies are overlapped with the accelerator's compute, as compared to CPU execution where Input Gather and Output Write incur high sequential latency.
In the middle layers, reuse opportunity is significantly higher, therefore the impact of dataflow optimization is low, yet speedups close to 20x are achieved over 1-CPU.
AccSS3D achieves a power reduction by \convprsc and \convprfc on average across all layers compared to 1-CPU and 4-CPU respectively.
Since middle layers are convolution heavy, CPU achieves higher instructions-per-cycle resulting in higher power consumption. Therefore, power reduction with AccSS3D is relatively higher for the middle layers.

\begin{figure}
	\centering
	\includegraphics[width=0.47\textwidth]{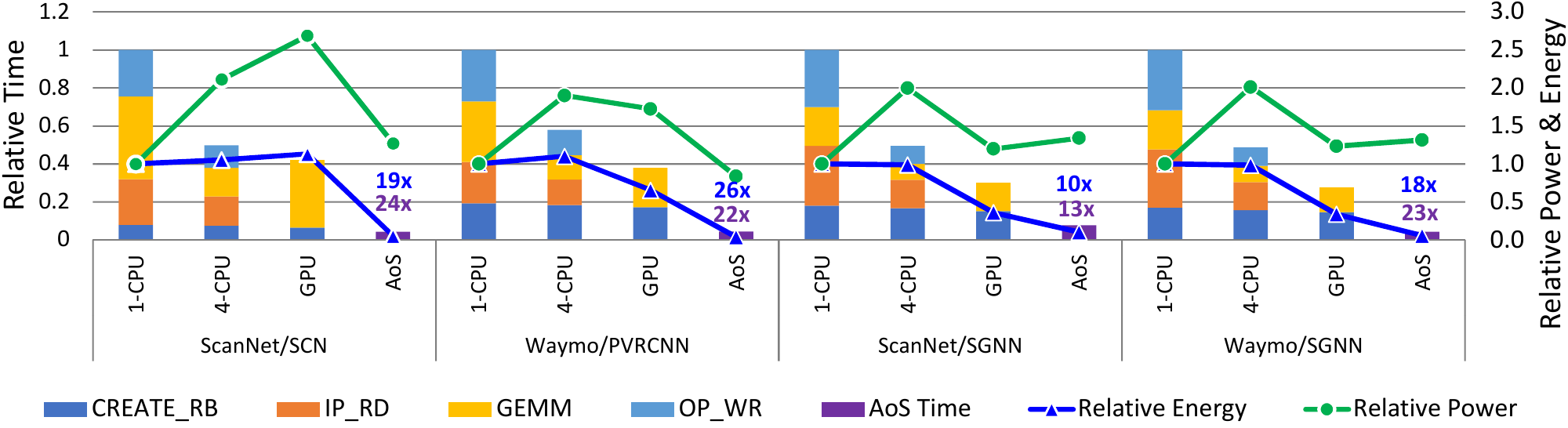}
    \caption{Power, Performance and Energy Savings (including DRAM) across workloads and datasets with AccSS3D over CPU and GPU baselines.}
	\label{fig:perf_network_dataset}
\end{figure}

Table \ref{tab:scn_savings} summarises speed-up and energy reduction for both 3D sparse convolution operation and end-to-end scene segmentation over CPU baselines for SCN on ScanNet.
Figure \ref{fig:perf_network_dataset} shows power and performance for two additional workloads (PV-RCNN, SGNN) and including Waymo's outdoor dataset, where AccSS3D is compared with 1-CPU, 4-CPU and GPU.
For PV-RCNN and SGNN networks, acceleration for \textit{Adjacency Map} creation provides significant speed-up and energy savings as size of pointcloud is relatively higher than number of channels ($N, C$) in these network topologies. 

\subsection{AccSS3D Feature Analysis}
\begin{figure}
	\centering
	\includegraphics[width=0.5\textwidth]{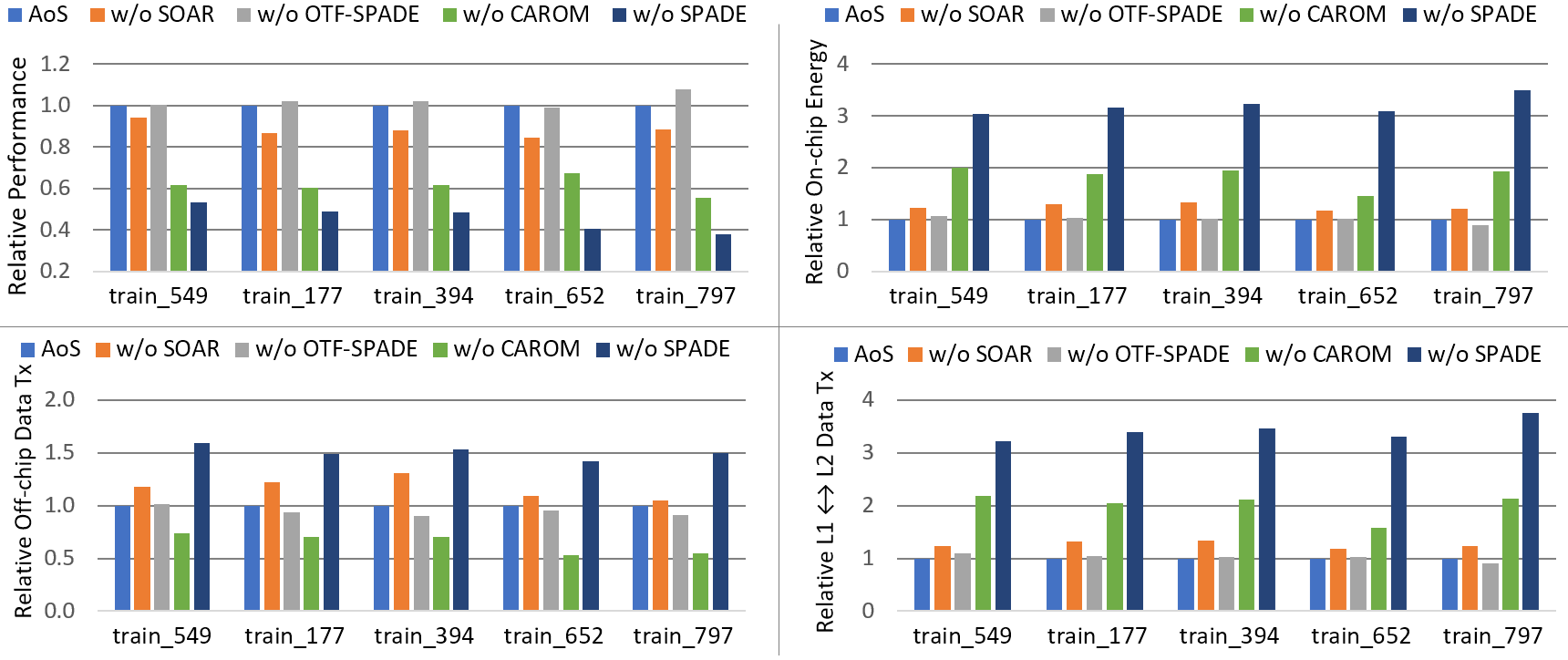}
	\caption{Impact of different AccSS3D features (\soar, \spade, \carom, \otfspade) on performance, on-chip energy savings, and DRAM data accesses evaluated across different input pointclouds in ScanNet dataset}. 
	\label{fig:feature_perf}
\end{figure}

\begin{figure}
  \centering
  \includegraphics[width=0.45\textwidth]{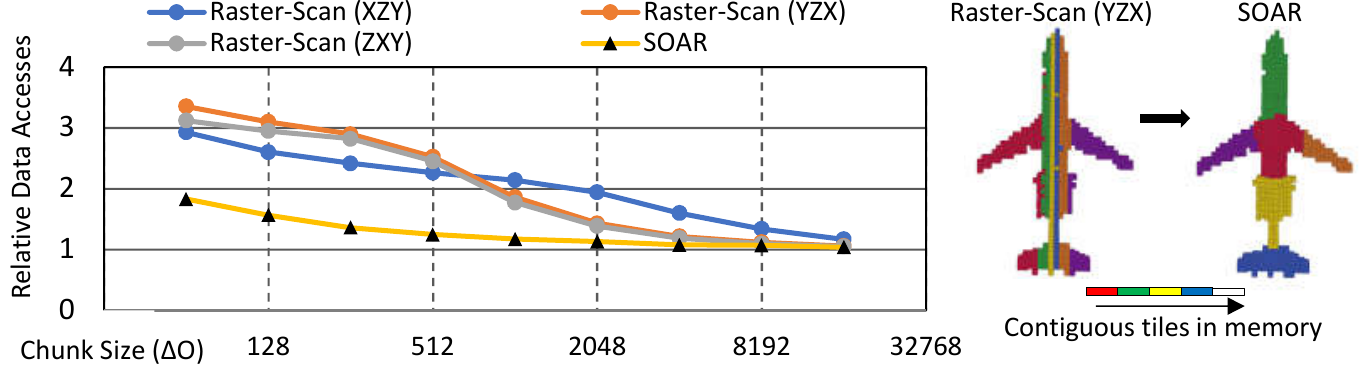}
  \caption{Left: Data access savings with \soar reordering over raster-scan \{inner, middle, outer loop\}. %
  Right: Effect of SOAR on a pointcloud; the thin strips caused by raster scan would invoke multiple data refetches from neighbouring tiles during convolution.}
  \label{fig:data_loc_with_soar}
\end{figure}

We evaluate the goodness of AccSS3D features as described in Sections \ref{sec:acc_for_ss} and \ref{sec:arch} (Figure \ref{fig:feature_perf}).
We collect performance metrics for each feature by disabling it from the fully-featured AccSS3D.
As reference software implements a weight-stationary dataflow with \textit{no spatial tiling}, it was infeasible to implement for few layers due to large weight size and also likely to be unoptimized for a system with limited memory such as the proposed AccSS3D.
Hence, we picked input-stationary dataflow as a reasonable baseline which performs tiling on available on-chip memory, equally distributing tiles along output channels ($N$) onto many-cores and supporting on-chip partial accumulation minimizing DRAM accesses.
\textbf{Optimal tiling and walk-pattern selected by \spade} provides significant reduction in both on-chip and DRAM data accesses compared to baseline dataflow.
A data-accesses minimization at L2 results in lower DRAM accesses, but it increases on-chip data transfers and hence performance drops significantly due to on-chip bandwidth bottlenecks.
\textbf{\carom} helps alleviating this issue by striking a balance between the DRAM and the on-chip data-transfers without being bandwidth limited at DRAM interface.
Comparing with input pointcloud dependent $\mathcal{SA}$ based dataflow ($\mathscr{ISA}$), \textbf{AccSS3D with $\mathscr{MSA}$} marginally loses performance for a few pointclouds while it provides significant reduction in end-to-end latency through offline-dataflow.
Gains with \textbf{\soar} based pointcloud reordering varies across pointclouds as scope for the reordering depends on input geometry and scan order performed during the data acquisition. Figure \ref{fig:data_loc_with_soar} shows relative data-access savings with \soar comparing with three different scan-orders along x, y and z direction.

\subsection{CPU Performance with \spade}
\begin{figure}
  \centering
  \includegraphics[width=0.5\textwidth]{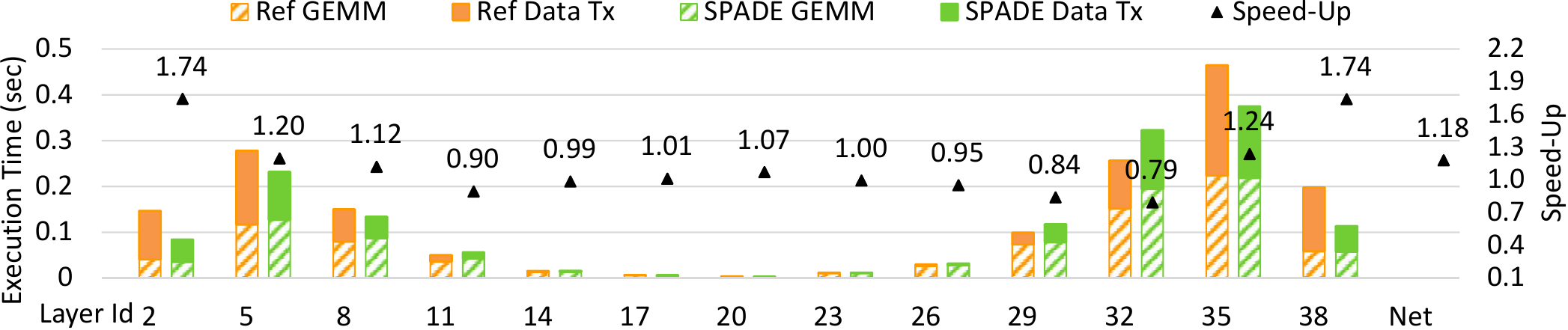}
  \caption{CPU performance improvements with \spade}
  \label{fig:spade_sw_speedup}
\end{figure}

To evaluate the performance impact of \spade on a CPU only system, we implement tiling and loop order in the reference SCN-CPU baseline \cite{graham20183d}.
Performing the 3D sparse convolution as per \coir metadata structure requires irregular data accesses interleaved with compute.
Without explicitly ensuring data residency in inner level caches, processing performance will be limited by latency.
For efficient processing, irregular data accesses need to be separated from the convolution.
To achieve this, similar to the reference CPU baseline, we gather input and weights into local buffers and after convolution we perform scattered write for output features. 
Since footprint for most of layers exceeds the capacity of the CPU's last-level-cache (LLC) and given the high latency to DRAM memory, we optimize for DRAM accesses.
The dataflow optimizer assumes 10\% of LLC's capacity to be used for code and miscellaneous data and remaining 90\% would be be available for the SCN's working set.
Figure \ref{fig:spade_sw_speedup} shows the CPU performance with the optimized tiling and loop order as recommended by \spade.
For brevity, we show one layer from each spatial resolution.
With \spade, overall performance improves by 18\%. 
For some layers gain performance upto 74\%, while for a few layers it drops by 21\%.
These layers have smaller metadata (due to lower resolution) and more channels ($C, N$), hence \spade prefers tiling across channels requiring metadata to be accessed and processed repeatedly.
\bfit{We observe that high CPU overheads of metadata processing and explicit scatter-gather operation are reasons for lower performance}.

\section{Related Work}
Taichi \cite{hu2019taichi} offers a high level interface to efficient data structures for spatially sparse data by using index analysis. %
Eyeriss \cite{chen2016eyeriss} proposed row-stationary dataflow with diagonal data-feedforwarding over 2D systolic array demonstrating substantial energy savings, while Eyeriss-V2 \cite{chen2018eyeriss} extended it to a scale-up architecture through mesh connections.
\cite{yang2016systematic} utilizes a tiling structure through unrolling nested loops of convolution to maximize reuse at different levels of caches.
FlexFlow \cite{lu2017flexflow} aims to maximize compute utilization by minimizing wastage due to spatial split of work among PEs.
Morph \cite{hegde2018morph} optimizes dataflows for a 3-level memory scale-up architecture maximizing reuse at each level in the hierarchy.
ExTensor \cite{hegde2019extensor} proposes technique to find non-zero element intersection for effective computation.
Sparten \cite{gondimalla2019sparten} defines SparseMap, which is two tuple of bitmask and performs efficient inner join logic to feed MAC.
SMAASH \cite{Smash} compresses sparse matrix in software and performs efficient index finding by hardware accelerator on compressed data.
\cite{HuaFineGrainedSparsity} proposes fine grained channel gating technique and an accelerator to exploit the dynamic sparsity.
\cite{zhang2016cambricon} proposes an accelerator with Indexing Module to efficiently select and transfer needed neurons to PEs.

\section{Conclusion}
Understanding of 3D objects and environment is critical for many real world applications.
To our best of knowledge, this is the first end-to-end solution for accelerating 3D scene analysis by exploiting spatial sparsity through sparsity-aware dataflow optimizer, novel micro-architecture and design for a spatially-sparse compute engine and, employing custom software-hardware co-designed methodologies.

\bibliographystyle{IEEEtranS}
\bibliography{spade}

% Generated by IEEEtranS.bst, version: 1.13 (2008/09/30)
\begin{thebibliography}{10}
\providecommand{\url}[1]{#1}
\csname url@samestyle\endcsname
\providecommand{\newblock}{\relax}
\providecommand{\bibinfo}[2]{#2}
\providecommand{\BIBentrySTDinterwordspacing}{\spaceskip=0pt\relax}
\providecommand{\BIBentryALTinterwordstretchfactor}{4}
\providecommand{\BIBentryALTinterwordspacing}{\spaceskip=\fontdimen2\font plus
\BIBentryALTinterwordstretchfactor\fontdimen3\font minus
  \fontdimen4\font\relax}
\providecommand{\BIBforeignlanguage}[2]{{%
\expandafter\ifx\csname l@#1\endcsname\relax
\typeout{** WARNING: IEEEtranS.bst: No hyphenation pattern has been}%
\typeout{** loaded for the language `#1'. Using the pattern for}%
\typeout{** the default language instead.}%
\else
\language=\csname l@#1\endcsname
\fi
#2}}
\providecommand{\BIBdecl}{\relax}
\BIBdecl

\bibitem{synopsysdc}
\BIBentryALTinterwordspacing
``{Tool from Synopsys, Design Compiler (Version L-2016.03-SP2)}.'' [Online].
  Available: \url{https://www.synopsys.com}
\BIBentrySTDinterwordspacing

\bibitem{synopsysic}
\BIBentryALTinterwordspacing
``{Tool from Synopsys, IC Compiler II (Version L-2016.03-SP5)}.'' [Online].
  Available: \url{https://www.synopsys.com}
\BIBentrySTDinterwordspacing

\bibitem{amplifier2019intel}
\BIBentryALTinterwordspacing
I.~V. Amplifier, ``Intel vtune amplifier,'' 2019. [Online]. Available:
  \url{https://software.intel.com/en-us/vtune}
\BIBentrySTDinterwordspacing

\bibitem{ayachi2018strided}
R.~Ayachi, M.~Afif, Y.~Said, and M.~Atri, ``Strided convolution instead of max
  pooling for memory efficiency of convolutional neural networks,'' in
  \emph{International conference on the Sciences of Electronics, Technologies
  of Information and Telecommunications}.\hskip 1em plus 0.5em minus
  0.4em\relax Springer, 2018, pp. 234--243.

\bibitem{celis1985robin}
\BIBentryALTinterwordspacing
P.~Celis, P.-A. Larson, and J.~I. Munro, ``Robin hood hashing,'' in \emph{26th
  Annual Symposium on Foundations of Computer Science (sfcs 1985)}.\hskip 1em
  plus 0.5em minus 0.4em\relax IEEE, 1985, pp. 281--288. [Online]. Available:
  \url{https://github.com/martinus/robin-hood-hashing}
\BIBentrySTDinterwordspacing

\bibitem{chen2017multi}
X.~Chen, H.~Ma, J.~Wan, B.~Li, and T.~Xia, ``Multi-view 3d object detection
  network for autonomous driving,'' in \emph{Proceedings of the IEEE Conference
  on Computer Vision and Pattern Recognition}, 2017, pp. 1907--1915.

\bibitem{chen2016eyeriss}
Y.-H. Chen, J.~Emer, and V.~Sze, ``Eyeriss: A spatial architecture for
  energy-efficient dataflow for convolutional neural networks,'' in \emph{ACM
  SIGARCH Computer Architecture News}, vol.~44, no.~3.\hskip 1em plus 0.5em
  minus 0.4em\relax IEEE Press, 2016, pp. 367--379.

\bibitem{chen2018eyeriss}
Y.-H. Chen, T.-J. Yang, J.~Emer, and V.~Sze, ``Eyeriss v2: A flexible
  accelerator for emerging deep neural networks on mobile devices,''
  \emph{arXiv preprint arXiv:1807.07928}, 2018.

\bibitem{chiang2019unified}
H.-Y. Chiang, Y.-L. Lin, Y.-C. Liu, and W.~H. Hsu, ``A unified point-based
  framework for 3d segmentation,'' in \emph{2019 International Conference on 3D
  Vision (3DV)}.\hskip 1em plus 0.5em minus 0.4em\relax IEEE, 2019, pp.
  155--163.

\bibitem{choy20194d}
C.~Choy, J.~Gwak, and S.~Savarese, ``4d spatio-temporal convnets: Minkowski
  convolutional neural networks,'' \emph{arXiv preprint arXiv:1904.08755},
  2019.

\bibitem{cciccek20163d}
{\"O}.~{\c{C}}i{\c{c}}ek, A.~Abdulkadir, S.~S. Lienkamp, T.~Brox, and
  O.~Ronneberger, ``3d u-net: learning dense volumetric segmentation from
  sparse annotation,'' in \emph{International conference on medical image
  computing and computer-assisted intervention}.\hskip 1em plus 0.5em minus
  0.4em\relax Springer, 2016, pp. 424--432.

\bibitem{MinkowskiEngCode}
\BIBentryALTinterwordspacing
N.~Corporation, ``Minkowski engine.'' [Online]. Available:
  \url{https://github.com/NVIDIA/MinkowskiEngine}
\BIBentrySTDinterwordspacing

\bibitem{dai2017scannet}
A.~Dai, A.~X. Chang, M.~Savva, M.~Halber, T.~Funkhouser, and M.~Nie{\ss}ner,
  ``Scannet: Richly-annotated 3d reconstructions of indoor scenes,'' in
  \emph{Proc. Computer Vision and Pattern Recognition (CVPR), IEEE}, 2017.

\bibitem{dai2019sg}
A.~Dai, C.~Diller, and M.~Nie{\ss}ner, ``Sg-nn: Sparse generative neural
  networks for self-supervised scene completion of rgb-d scans,'' \emph{arXiv
  preprint arXiv:1912.00036}, 2019.

\bibitem{dai2018scancomplete}
A.~Dai, D.~Ritchie, M.~Bokeloh, S.~Reed, J.~Sturm, and M.~Nie{\ss}ner,
  ``Scancomplete: Large-scale scene completion and semantic segmentation for 3d
  scans,'' in \emph{Proceedings of the IEEE Conference on Computer Vision and
  Pattern Recognition}, 2018, pp. 4578--4587.

\bibitem{acorn}
X.~{Dong}, L.~{Liu}, P.~{Zhao}, G.~{Li}, J.~{Li}, X.~{Wang}, and X.~{Feng},
  ``Acorns: A framework for accelerating deep neural networks with input
  sparsity,'' in \emph{2019 28th International Conference on Parallel
  Architectures and Compilation Techniques (PACT)}, 2019, pp. 178--191.

\bibitem{gao2019tangram}
M.~Gao, X.~Yang, J.~Pu, M.~Horowitz, and C.~Kozyrakis, ``Tangram: Optimized
  coarse-grained dataflow for scalable nn accelerators,'' in \emph{Proceedings
  of the Twenty-Fourth International Conference on Architectural Support for
  Programming Languages and Operating Systems}.\hskip 1em plus 0.5em minus
  0.4em\relax ACM, 2019, pp. 807--820.

\bibitem{garbade2019two}
M.~Garbade, Y.-T. Chen, J.~Sawatzky, and J.~Gall, ``Two stream 3d semantic
  scene completion,'' in \emph{Proceedings of the IEEE Conference on Computer
  Vision and Pattern Recognition Workshops}, 2019, pp. 0--0.

\bibitem{gondimalla2019sparten}
A.~Gondimalla, N.~Chesnut, M.~Thottethodi, and T.~Vijaykumar, ``Sparten: A
  sparse tensor accelerator for convolutional neural networks,'' in
  \emph{Proceedings of the 52nd Annual IEEE/ACM International Symposium on
  Microarchitecture}.\hskip 1em plus 0.5em minus 0.4em\relax ACM, 2019, pp.
  151--165.

\bibitem{graham20183d}
B.~Graham, M.~Engelcke, and L.~van~der Maaten, ``3d semantic segmentation with
  submanifold sparse convolutional networks,'' in \emph{Proceedings of the IEEE
  Conference on Computer Vision and Pattern Recognition}, 2018, pp. 9224--9232.

\bibitem{graham2017submanifold}
B.~Graham and L.~van~der Maaten, ``Submanifold sparse convolutional networks,''
  \emph{arXiv preprint arXiv:1706.01307}, 2017.

\bibitem{guo2019deep}
Y.~Guo, H.~Wang, Q.~Hu, H.~Liu, L.~Liu, and M.~Bennamoun, ``Deep learning for
  3d point clouds: A survey,'' \emph{arXiv preprint arXiv:1912.12033}, 2019.

\bibitem{han2020occuseg}
L.~Han, T.~Zheng, L.~Xu, and L.~Fang, ``Occuseg: Occupancy-aware 3d instance
  segmentation,'' in \emph{Proceedings of the IEEE/CVF Conference on Computer
  Vision and Pattern Recognition}, 2020, pp. 2940--2949.

\bibitem{hegde2018morph}
K.~Hegde, R.~Agrawal, Y.~Yao, and C.~W. Fletcher, ``Morph: Flexible
  acceleration for 3d cnn-based video understanding,'' in \emph{2018 51st
  Annual IEEE/ACM International Symposium on Microarchitecture (MICRO)}.\hskip
  1em plus 0.5em minus 0.4em\relax IEEE, 2018, pp. 933--946.

\bibitem{hegde2019extensor}
K.~Hegde, H.~Asghari-Moghaddam, M.~Pellauer, N.~Crago, A.~Jaleel, E.~Solomonik,
  J.~Emer, and C.~W. Fletcher, ``Extensor: An accelerator for sparse tensor
  algebra,'' in \emph{Proceedings of the 52nd Annual IEEE/ACM International
  Symposium on Microarchitecture}.\hskip 1em plus 0.5em minus 0.4em\relax ACM,
  2019, pp. 319--333.

\bibitem{hu2019taichi}
Y.~Hu, T.-M. Li, L.~Anderson, J.~Ragan-Kelley, and F.~Durand, ``Taichi: a
  language for high-performance computation on spatially sparse data
  structures,'' \emph{ACM Transactions on Graphics (TOG)}, vol.~38, no.~6, pp.
  1--16, 2019.

\bibitem{HuaFineGrainedSparsity}
\BIBentryALTinterwordspacing
W.~Hua, Y.~Zhou, C.~De~Sa, Z.~Zhang, and G.~E. Suh, ``Boosting the performance
  of cnn accelerators with dynamic fine-grained channel gating,'' in
  \emph{Proceedings of the 52Nd Annual IEEE/ACM International Symposium on
  Microarchitecture}, ser. MICRO '52.\hskip 1em plus 0.5em minus 0.4em\relax
  New York, NY, USA: ACM, 2019, pp. 139--150. [Online]. Available:
  \url{http://doi.acm.org/10.1145/3352460.3358283}
\BIBentrySTDinterwordspacing

\bibitem{centaur}
R.~Hwang, T.~Kim, Y.~Kwon, and M.~Rhu, ``Centaur: A chiplet-based, hybrid
  sparse-dense accelerator for personalized recommendations,'' in \emph{2020
  ACM/IEEE 47th Annual International Symposium on Computer Architecture
  (ISCA)}.\hskip 1em plus 0.5em minus 0.4em\relax IEEE, 2020, p.~1.

\bibitem{googlehashmap}
\BIBentryALTinterwordspacing
G.~Inc, ``Google sparse hash.'' [Online]. Available:
  \url{https://github.com/sparsehash/sparsehash}
\BIBentrySTDinterwordspacing

\bibitem{ji20123d}
S.~Ji, W.~Xu, M.~Yang, and K.~Yu, ``3d convolutional neural networks for human
  action recognition,'' \emph{IEEE transactions on pattern analysis and machine
  intelligence}, vol.~35, no.~1, pp. 221--231, 2012.

\bibitem{jiang2019hierarchical}
L.~Jiang, H.~Zhao, S.~Liu, X.~Shen, C.-W. Fu, and J.~Jia, ``Hierarchical
  point-edge interaction network for point cloud semantic segmentation,'' in
  \emph{Proceedings of the IEEE International Conference on Computer Vision},
  2019, pp. 10\,433--10\,441.

\bibitem{jiang2018pointsift}
M.~Jiang, Y.~Wu, T.~Zhao, Z.~Zhao, and C.~Lu, ``Pointsift: A sift-like network
  module for 3d point cloud semantic segmentation,'' \emph{arXiv preprint
  arXiv:1807.00652}, 2018.

\bibitem{Smash}
\BIBentryALTinterwordspacing
K.~Kanellopoulos, N.~Vijaykumar, C.~Giannoula, R.~Azizi, S.~Koppula, N.~M.
  Ghiasi, T.~Shahroodi, J.~G. Luna, and O.~Mutlu, ``Smash: Co-designing
  software compression and hardware-accelerated indexing for efficient sparse
  matrix operations,'' in \emph{Proceedings of the 52Nd Annual IEEE/ACM
  International Symposium on Microarchitecture}, ser. MICRO '52.\hskip 1em plus
  0.5em minus 0.4em\relax New York, NY, USA: ACM, 2019, pp. 600--614. [Online].
  Available: \url{http://doi.acm.org/10.1145/3352460.3358286}
\BIBentrySTDinterwordspacing

\bibitem{taco}
F.~Kjolstad, S.~Chou, D.~Lugato, S.~Kamil, and S.~Amarasinghe, ``taco: A tool
  to generate tensor algebra kernels,'' in \emph{2017 32nd IEEE/ACM
  International Conference on Automated Software Engineering (ASE)}, Oct 2017,
  pp. 943--948.

\bibitem{kwon2018mef}
H.~Kwon, A.~Samajdar, and T.~Krishna, ``Maeri: Enabling flexible dataflow
  mapping over dnn accelerators via reconfigurable interconnects,'' in
  \emph{Proceedings of the Twenty-Third International Conference on
  Architectural Support for Programming Languages and Operating Systems}, ser.
  ASPLOS '18.\hskip 1em plus 0.5em minus 0.4em\relax New York, NY, USA: ACM,
  2018, pp. 461--475.

\bibitem{li2019rgbd}
J.~Li, Y.~Liu, D.~Gong, Q.~Shi, X.~Yuan, C.~Zhao, and I.~Reid, ``Rgbd based
  dimensional decomposition residual network for 3d semantic scene
  completion,'' in \emph{Proceedings of the IEEE Conference on Computer Vision
  and Pattern Recognition}, 2019, pp. 7693--7702.

\bibitem{li2019three}
X.~Li, J.~Guivant, N.~Kwok, Y.~Xu, R.~Li, and H.~Wu, ``Three-dimensional
  backbone network for 3d object detection in traffic scenes,'' \emph{arXiv
  preprint arXiv:1901.08373}, 2019.

\bibitem{li2018pointcnn}
Y.~Li, R.~Bu, M.~Sun, W.~Wu, X.~Di, and B.~Chen, ``Pointcnn: Convolution on
  x-transformed points,'' in \emph{Advances in neural information processing
  systems}, 2018, pp. 820--830.

\bibitem{li2019pointsite}
Z.~Li, X.~Yan, Q.~Wei, X.~Gao, S.~Wang, and S.~Cui, ``Pointsite: a point cloud
  segmentation tool for identication of protein ligand binding atoms,''
  \emph{bioRxiv}, p. 831131, 2019.

\bibitem{liang20203d}
Z.~Liang, M.~Yang, L.~Hao, and C.~Wang, ``3d instance embedding learning with a
  structure-aware loss function for point cloud segmentation,'' \emph{IEEE
  Robotics and Automation Letters}, 2020.

\bibitem{liu2018see}
S.~Liu, Y.~Hu, Y.~Zeng, Q.~Tang, B.~Jin, Y.~Han, and X.~Li, ``See and think:
  Disentangling semantic scene completion,'' in \emph{Advances in Neural
  Information Processing Systems}, 2018, pp. 263--274.

\bibitem{lu2017flexflow}
W.~Lu, G.~Yan, J.~Li, S.~Gong, Y.~Han, and X.~Li, ``Flexflow: A flexible
  dataflow accelerator architecture for convolutional neural networks,'' in
  \emph{2017 IEEE International Symposium on High Performance Computer
  Architecture (HPCA)}.\hskip 1em plus 0.5em minus 0.4em\relax IEEE, 2017, pp.
  553--564.

\bibitem{maturana2015voxnet}
D.~Maturana and S.~Scherer, ``Voxnet: A 3d convolutional neural network for
  real-time object recognition,'' in \emph{2015 IEEE/RSJ International
  Conference on Intelligent Robots and Systems (IROS)}.\hskip 1em plus 0.5em
  minus 0.4em\relax IEEE, 2015, pp. 922--928.

\bibitem{micron2020power}
\BIBentryALTinterwordspacing
D.~Micron, ``Power calculator,'' 2020. [Online]. Available:
  \url{https://www.micron.com/support/tools-and-utilities/power-calc}
\BIBentrySTDinterwordspacing

\bibitem{mubeen2018workload}
\BIBentryALTinterwordspacing
N.~Mubeen, ``Workload frequency scaling law-derivation and verification,''
  \emph{Queue}, vol.~16, no.~2, pp. 50--66, 2018. [Online]. Available:
  \url{https://github.com/intel/psst}
\BIBentrySTDinterwordspacing

\bibitem{nvidia2016nvidia}
\BIBentryALTinterwordspacing
Nvidia, ``Nvidia visual profiler,'' 2016. [Online]. Available:
  \url{https://developer.nvidia.com/nvidia-visual-profiler}
\BIBentrySTDinterwordspacing

\bibitem{parashar2017scnn}
A.~Parashar, M.~Rhu, A.~Mukkara, A.~Puglielli, R.~Venkatesan, B.~Khailany,
  J.~Emer, S.~W. Keckler, and W.~J. Dally, ``Scnn: An accelerator for
  compressed-sparse convolutional neural networks,'' in \emph{2017 ACM/IEEE
  44th Annual International Symposium on Computer Architecture (ISCA)}.\hskip
  1em plus 0.5em minus 0.4em\relax IEEE, 2017, pp. 27--40.

\bibitem{pham2019real}
Q.-H. Pham, B.-S. Hua, T.~Nguyen, and S.-K. Yeung, ``Real-time progressive 3d
  semantic segmentation for indoor scenes,'' in \emph{2019 IEEE Winter
  Conference on Applications of Computer Vision (WACV)}.\hskip 1em plus 0.5em
  minus 0.4em\relax IEEE, 2019, pp. 1089--1098.

\bibitem{qi2017pointnet}
C.~R. Qi, H.~Su, K.~Mo, and L.~J. Guibas, ``Pointnet: Deep learning on point
  sets for 3d classification and segmentation,'' in \emph{Proceedings of the
  IEEE Conference on Computer Vision and Pattern Recognition}, 2017, pp.
  652--660.

\bibitem{qi2017pointnet++}
C.~R. Qi, L.~Yi, H.~Su, and L.~J. Guibas, ``Pointnet++: Deep hierarchical
  feature learning on point sets in a metric space,'' in \emph{Advances in
  neural information processing systems}, 2017, pp. 5099--5108.

\bibitem{SparseConvCode}
\BIBentryALTinterwordspacing
F.~Research, ``Submanifold sparse convolutional.'' [Online]. Available:
  \url{https://github.com/facebookresearch/SparseConvNet/tree/master/sparseconvnet/SCN}
\BIBentrySTDinterwordspacing

\bibitem{rosu2019latticenet}
R.~A. Rosu, P.~Sch{\"u}tt, J.~Quenzel, and S.~Behnke, ``Latticenet: Fast point
  cloud segmentation using permutohedral lattices,'' \emph{arXiv preprint
  arXiv:1912.05905}, 2019.

\bibitem{schmohl2019submanifold}
S.~Schmohl and U.~S{\"o}rgel, ``Submanifold sparse convolutional networks for
  semantic segmentation of large-scale als point clouds.'' \emph{ISPRS Annals
  of Photogrammetry, Remote Sensing \& Spatial Information Sciences}, vol.~4,
  2019.

\bibitem{shi2019pv}
S.~Shi, C.~Guo, L.~Jiang, Z.~Wang, J.~Shi, X.~Wang, and H.~Li, ``Pv-rcnn:
  Point-voxel feature set abstraction for 3d object detection,'' \emph{arXiv
  preprint arXiv:1912.13192}, 2019.

\bibitem{shi2020points}
S.~Shi, Z.~Wang, J.~Shi, X.~Wang, and H.~Li, ``From points to parts: 3d object
  detection from point cloud with part-aware and part-aggregation network,''
  \emph{IEEE Transactions on Pattern Analysis and Machine Intelligence}, 2020.

\bibitem{sun2019scalability}
P.~Sun, H.~Kretzschmar, X.~Dotiwalla, A.~Chouard, V.~Patnaik, P.~Tsui, J.~Guo,
  Y.~Zhou, Y.~Chai, B.~Caine \emph{et~al.}, ``Scalability in perception for
  autonomous driving: Waymo open dataset,'' \emph{arXiv}, pp. arXiv--1912,
  2019.

\bibitem{tatarchenko2018tangent}
M.~Tatarchenko, J.~Park, V.~Koltun, and Q.-Y. Zhou, ``Tangent convolutions for
  dense prediction in 3d,'' in \emph{Proceedings of the IEEE Conference on
  Computer Vision and Pattern Recognition}, 2018, pp. 3887--3896.

\bibitem{teschner2003optimized}
M.~Teschner, B.~Heidelberger, M.~M{\"u}ller, D.~Pomerantes, and M.~H. Gross,
  ``Optimized spatial hashing for collision detection of deformable objects.''
  in \emph{Vmv}, vol.~3, 2003, pp. 47--54.

\bibitem{thomas2019kpconv}
H.~Thomas, C.~R. Qi, J.-E. Deschaud, B.~Marcotegui, F.~Goulette, and L.~J.
  Guibas, ``Kpconv: Flexible and deformable convolution for point clouds,'' in
  \emph{Proceedings of the IEEE International Conference on Computer Vision},
  2019, pp. 6411--6420.

\bibitem{graphcore}
\BIBentryALTinterwordspacing
N.~Toon and S.~Knowles, ``Graphcore,'' 2017. [Online]. Available:
  \url{https://www.graphcore.ai}
\BIBentrySTDinterwordspacing

\bibitem{tran2015learning}
D.~Tran, L.~Bourdev, R.~Fergus, L.~Torresani, and M.~Paluri, ``Learning
  spatiotemporal features with 3d convolutional networks,'' in
  \emph{Proceedings of the IEEE international conference on computer vision},
  2015, pp. 4489--4497.

\bibitem{uckermann2012real}
A.~{\"U}ckermann, R.~Haschke, and H.~Ritter, ``Real-time 3d segmentation of
  cluttered scenes for robot grasping,'' in \emph{2012 12th IEEE-RAS
  International Conference on Humanoid Robots (Humanoids 2012)}.\hskip 1em plus
  0.5em minus 0.4em\relax IEEE, 2012, pp. 198--203.

\bibitem{wang20203d}
L.~Wang, X.~Fan, J.~Chen, J.~Cheng, J.~Tan, and X.~Ma, ``3d object detection
  based on sparse convolution neural network and feature fusion for autonomous
  driving in smart cities,'' \emph{Sustainable Cities and Society}, vol.~54, p.
  102002, 2020.

\bibitem{wu20153d}
Z.~Wu, S.~Song, A.~Khosla, F.~Yu, L.~Zhang, X.~Tang, and J.~Xiao, ``3d
  shapenets: A deep representation for volumetric shapes,'' in
  \emph{Proceedings of the IEEE conference on computer vision and pattern
  recognition}, 2015, pp. 1912--1920.

\bibitem{Tigris}
\BIBentryALTinterwordspacing
T.~Xu, B.~Tian, and Y.~Zhu, ``Tigris: Architecture and algorithms for 3d
  perception in point clouds,'' in \emph{Proceedings of the 52Nd Annual
  IEEE/ACM International Symposium on Microarchitecture}, ser. MICRO '52.\hskip
  1em plus 0.5em minus 0.4em\relax New York, NY, USA: ACM, 2019, pp. 629--642.
  [Online]. Available: \url{http://doi.acm.org/10.1145/3352460.3358259}
\BIBentrySTDinterwordspacing

\bibitem{yang2016systematic}
X.~Yang, J.~Pu, B.~B. Rister, N.~Bhagdikar, S.~Richardson, S.~Kvatinsky,
  J.~Ragan-Kelley, A.~Pedram, and M.~Horowitz, ``A systematic approach to
  blocking convolutional neural networks,'' \emph{arXiv preprint
  arXiv:1606.04209}, 2016.

\bibitem{ye2020sarpnet}
Y.~Ye, H.~Chen, C.~Zhang, X.~Hao, and Z.~Zhang, ``Sarpnet: Shape attention
  regional proposal network for lidar-based 3d object detection,''
  \emph{Neurocomputing}, vol. 379, pp. 53--63, 2020.

\bibitem{ye2019arpnet}
Y.~Ye, C.~Zhang, and X.~Hao, ``Arpnet: attention region proposal network for 3d
  object detection,'' \emph{Science China Information Sciences}, vol.~62,
  no.~12, p. 220104, 2019.

\bibitem{zhang2018efficient}
J.~Zhang, H.~Zhao, A.~Yao, Y.~Chen, L.~Zhang, and H.~Liao, ``Efficient semantic
  scene completion network with spatial group convolution,'' in
  \emph{Proceedings of the European Conference on Computer Vision (ECCV)},
  2018, pp. 733--749.

\bibitem{zhang2019cascaded}
P.~Zhang, W.~Liu, Y.~Lei, H.~Lu, and X.~Yang, ``Cascaded context pyramid for
  full-resolution 3d semantic scene completion,'' in \emph{Proceedings of the
  IEEE International Conference on Computer Vision}, 2019, pp. 7801--7810.

\bibitem{zhang2016cambricon}
S.~Zhang, Z.~Du, L.~Zhang, H.~Lan, S.~Liu, L.~Li, Q.~Guo, T.~Chen, and Y.~Chen,
  ``Cambricon-x: An accelerator for sparse neural networks,'' in \emph{The 49th
  Annual IEEE/ACM International Symposium on Microarchitecture}.\hskip 1em plus
  0.5em minus 0.4em\relax IEEE Press, 2016, p.~20.

\end{thebibliography}

\end{document}